\documentclass[aip,reprint]{revtex4-1}
\usepackage{graphicx}
\usepackage{amsmath, cases}
\usepackage{dcolumn}
\usepackage{bm}
\usepackage{ulem}
\usepackage{times}
\usepackage{amssymb}
\usepackage{epsfig}
\usepackage{xcolor}
\usepackage[T1]{fontenc}
\usepackage{mathtools, cases}   

\draft 

\begin{document}
\title{Generalized mode-coupling theory of the glass transition. II. Analytical scaling laws}
\author{Chengjie Luo}
\email[Electronic mail: ]{C.Luo@tue.nl}
\affiliation{Theory of Polymers and Soft Matter, Department of Applied Physics,
	Eindhoven University of Technology, P.O. Box 513, 5600MB Eindhoven, The Netherlands}
\author{Liesbeth M.~C.~Janssen}
\email[Electronic mail: ]{L.M.C.Janssen@tue.nl}
\affiliation{Theory of Polymers and Soft Matter, Department of Applied Physics,
	Eindhoven University of Technology, P.O. Box 513, 5600MB Eindhoven, The Netherlands}

\date{\today}

\begin{abstract}
	Generalized mode-coupling theory (GMCT) constitutes a systematically correctable, first-principles theory to study the dynamics of supercooled liquids and the glass transition. It is a hierarchical framework that, through the incorporation of increasingly many particle density correlations, can remedy some of the inherent limitations of the ideal mode-coupling theory (MCT). However, despite MCT's limitations, the ideal theory also enjoys several remarkable successes, notably including the analytical scaling laws for the $\alpha$- and $\beta$-relaxation dynamics.  Here we mathematically derive  similar scaling laws for arbitrary-order multi-point density correlation functions obtained from GMCT under arbitrary mean-field closure levels. More specifically, we analytically derive the asymptotic and preasymptotic solutions for the long-time limits of multi-point density correlators, the critical dynamics with two power-law decays, the factorization scaling laws in the $\beta$-relaxation regime, and the time-density superposition principle in the $\alpha$-relaxation regime.  The two characteristic power-law-divergent relaxation times for the two-step decay and the non-trivial relation between their exponents are also obtained. The validity ranges of the leading-order scaling laws are also provided by considering  the leading preasymptotic corrections. Furthermore, we test these solutions for the Percus-Yevick hard-sphere system. We demonstrate that GMCT preserves all the celebrated scaling laws of MCT while quantitatively improving the exponents, rendering the theory a promising candidate for an ultimately quantitative first-principles theory of glassy dynamics. 
	 
\end{abstract}
\maketitle

%

The glass transition in supercooled liquids and dense colloidal suspensions poses a notoriously difficult problem in condensed matter science.\cite{debenedetti2001supercooled,berthier2011theoretical}  One of the key challenges is to understand how supercooling or compression of a glass-forming material can lead to a dramatic slowdown of the relaxation dynamics, while the microstructure, as seen in e.g.\ static two-point correlation functions, remains very close to that of a normal liquid. Within the broad pallet of theories put forward to rationalize the glass transition\cite{berthier2011theoretical,kirkpatrick2015colloquium,ritort2003glassy,
adam1965temperature, royall2015role, ediger2000spatially,tarjus2011overview,biroli2013perspective}---many of which contain phenomenological elements---, 
mode-coupling theory (MCT) \cite{gotze2008complex,leutheusser1984dynamical,bengtzelius1984dynamics, 
reichman2005mode,janssen2018mode} takes a unique approach by starting from the formally exact, microscopic picture of a correlated (supercooled) liquid.  In particular, using only the static structure as input, MCT can reproduce some of the most important features of the dynamics, which makes it essentially the only theory of glassy dynamics based purely on first principles.

In MCT, the microscopic dynamics is characterized by the two-point density correlation function $F(k,t)$, a microscopic quantity that can be measured in both experiment and simulation at a certain wavenumber $k$ and time $t$. The equation of the motion for $F(k,t)$ is controlled by a memory function, in which four-point density correlation functions are the leading terms. These four-point density correlation functions are usually approximated by the product of two $F(k,t)$'s at different wavenumbers, resulting in the wavenumber-coupled self-consistent equations of standard MCT. 

Despite the factorization approximation, the predictions of MCT are remarkably successful for several non-trivial features of glass formation. The first success, from a historical point of view, is the prediction of a two-step decay of $F(k,t)$ via the so-called $\beta$ and $\alpha$ relaxation processes, respectively, where the dramatic slowdown emerges as an increasingly long plateau of $F(k,t)$ in the $\beta$ regime at low temperatures or high densities. MCT also offers an intuitive explanation for this plateau in terms of the cage effect, which constitutes a non-linear dynamic feedback mechanism due to local particle crowding. The more subtle achievements include several scaling laws which are applicable for many glass-forming materials \cite{reichman2005mode, gotze2008complex, weysser2010structural, sciortino2001debye, horbach2001relaxation, berthier2010critical}. In particular, MCT predicts (i) that the two time scales $\tau_\beta$ and $\tau$ associated with the $\beta$ and $\alpha$ regimes  diverge as a power law of the reduced temperature (or the reduced packing fraction) with exponents $1/2a$ and $\gamma$ respectively; 
(ii) that a time-temperature or time-density superposition principle holds in the $\alpha$-relaxation regime;
and 
(iii) that there is a universal decay $g_{\pm}(t)$ in the $\beta$-relaxation regime after scaling with the wavenumber, the reduced temperature (or the reduced packing fraction), and the $\beta$-relaxation time $\tau_\beta$. More specifically, to leading order, the above $g_{+}(t)$ and $g_{-}(t)$ are power functions of $t$ with exponents $a$ and $-b$, respectively. These functions are also known as the critical decay and the von Schweidler law. Remarkably, within MCT the above exponents are related to each other via $\lambda=\Gamma(1-a)^2/\Gamma(1-2a)=\Gamma(1+b)^2/\Gamma(1+2b)$ and $\gamma=1/2a+1/2b$, where $\lambda$ is determined by the static structure factors at the glass transition point predicted by MCT. This constitutes a highly non-trivial connection between the early $\beta$, late $\beta$, and $\alpha$ process. All the above scaling laws as well as their validity ranges from the leading corrections were already analytically derived decades ago.\cite{gotze1985properties, Gotze1990,franosch1997asymptotic,gotze2008complex} Moreover, the $\alpha$-relaxation process predicted by MCT can be also well described by a stretched-exponential Kohlrausch function, in general agreement with experiment, and the stretching exponents are also related to the exponent $b$ at large wavenumbers.\cite{fuchs1994kohlrausch}  Because of the widely affirmative tests of the above analytical scaling laws, MCT undoubtedly catches some of the key information for the dynamics of glass formation, although not all glass-forming materials fulfill all the above scaling laws. 

However, there are also several limitations of MCT. First of all, MCT typically overestimates (underestimates) the critical temperature $T_g$ (density $\varphi_g$) of the glass transition.\cite{reichman2005mode} Therefore, nearly all the theoretical analyses of the above scaling laws for $F(k,t)$ have to be done after rescaling the critical point with respect to the experimental glass transition, in order to admit a meaningful comparison with experimental or simulation results. Another failure is on the prediction of the fragility, i.e.\ the abruptness of a glass-forming material transitioning from liquid to glass.\cite{tarjus2014assessment} MCT always predicts a power-law divergence of the $\alpha$-relaxation time or viscosity, which may account for fragile systems but not for strong systems such as silica\cite{berthier2011theoretical}, and more generally disagrees with the empirical Vogel-Fulcher-Tamman (VFT) law. In general, MCT can be regarded as a type of mean-field theory\cite{kim2014equilibrium} which neglects activated dynamics, thus one may expect the theory to be only applicable in the mildly supercooled regime; this limitation may partly explain the above failures. Moreover, this mean-field theory is not consistent with the one from a thermodynamic point of view in the high-dimensional limit.\cite{ikeda2010mode,schmid2010glass,maimbourg2016solution} The inherent lack of activated dynamics also prevents MCT, in its standard form, to account for dynamical heterogeneity\cite{biroli2006inhomogeneous} and the violation of Stokes-Einstein relation in deeply supercooled liquids. 

To solve the above problems, a generalized mode-coupling theory (GMCT) was proposed by Szamel\cite{szamel2003colloidal} and developed further in recent years.\cite{ wu2005high,mayer2006cooperativity,janssen2014relaxation,janssen2015microscopic,
janssen2016generalized} 
The difference between MCT and GMCT starts from the approximation of the four-point density correlation functions. Instead of being factorized into the product of two $F(k,t)$'s, the dynamics of the four-point density correlation functions is described by a new exact equation with a new memory function, in which six-point density correlation functions are the leading terms. This procedure can be continued and finally a hierarchy of coupled integro-differential equations can be obtained. In this way, the uncontrolled approximation, i.e.\ the factorization of high-order density correlation functions to the product of lower orders, is postponed to arbitrary high orders to close the equations, or even strictly avoided when the order goes to infinity. A rigorous mathematical analysis has also confirmed existence and uniqueness of solutions to such GMCT hierarchies at arbitrary finite order.\cite{biezemans2020glassy}

Recent studies showed that GMCT can indeed improve the prediction of $\varphi^c$ for glassy hard spheres.\cite{szamel2003colloidal,wu2005high} A more recent study on weakly polydisperse hard spheres also showed that the time-dependent two-point density correlation functions converge to the simulation data when the orders included in GMCT increase.\cite{janssen2015microscopic} In our accompanying paper,\cite{luo2019generalized} we numerically test that all the above scaling laws for $F(k,t)$ in MCT are still applicable in GMCT for Percus-Yevick hard spheres. More intriguingly, we find that the  exponents characterizing the scaling laws ($\lambda,a,b,\gamma$ and so on) are quantitatively improved.
Notably, the improved $\gamma$, which specifies the $\alpha$-relaxation behavior, demonstrates that GMCT is also able to tune the degree of fragility with increasing order. This capability of accounting for different degrees of fragility was also studied in several wavevector-independent schematic GMCT models, in which it was found that both fragile and strong relaxations can emerge within infinite GMCT hierarchies.\cite{janssen2014relaxation} Overall, these results indicate that GMCT is a promising first-principles-based approach to extend the applicability range of MCT-like methods qualitatively and quantitatively.

At the moment, however, we still lack a full understanding of the dynamical effect of the hierarchical equations of multi-point density correlation functions in microscopic GMCT. On the one hand, all reports on microscopic GMCT calculations thus far have only numerically tested the quantitative improvement of $F(k,t)$ and the preserved scaling laws, but neither the mathematical explanations nor the applicability range of the scaling laws have been provided. Therefore, for GMCT, asymptotic results with leading corrections of $F(k,t)$ similar to those in MCT are necessary. On the other hand, previous studies have mainly focused on the prediction of the two-point density functions, while the dynamics of the higher-order density correlation functions has not yet been rigorously studied within the GMCT framework. In particular, the behavior of the four-point density correlation functions, which are related to dynamical heterogeneity, are vital for checking the possible underlying activated dynamics of glass formation. Hence the dynamics of multi-point density correlation functions as well as their asymptotic laws are also desired.

In this paper, we show the derivation of the asymptotic laws and some of the leading preasymptotic corrections of multi-point density correlation functions in the framework of GMCT. Following the accompanying paper,\cite{luo2019generalized} we use the Percus-Yevick (PY) hard sphere system\cite{wertheim1963exact} as a model to test our results when necessary. The asymptotic laws and the leading corrections for the standard MCT of PY hard spheres have been carefully studied in Ref.\ \onlinecite{gotze1985properties} and \onlinecite{franosch1997asymptotic}, and they provide the inspiration for this work. In the following, we first introduce the GMCT framework where the hierarchy of equations is provided. Then we present the long-time limit solutions for arbitrary-order density correlation functions in the vicinity of the critical point, in which the expansion technique for all scaling laws is introduced. Next we derive the dynamics of the correlators at the critical point, where the von Schweidler law naturally emerges.  Furthermore, the scaling laws in the $\beta$-relaxation regime for small reduced packing fractions are presented. The mentioned power law of the two relaxation times, the power decay $g_{\pm}(t)$, as well as the relations of the exponents are mathematically demonstrated. Finally, we establish the time-density (or time-temperature) superposition principle in the $\alpha$-relaxation regime, which is applicable in a wider density (or temperature) range than the scaling laws in the $\beta$-relaxation regime.

\section{GMCT equations}
We first summarize the microscopic GMCT equations of motion first derived in Ref.~\onlinecite{janssen2015microscopic} and also introduced in the accompanying paper.\cite{luo2019generalized} Within GMCT, the microscopic dynamics of a structural glass-former is described in terms of the normalized $2n$-point density correlation functions $\phi_n(k_1,\hdots,k_n,t)$, defined as 
\begin{equation}
\label{eq:phindef}
\phi_n(k_1,\hdots,k_n, t) = \frac{\langle \rho_{\bm{-k_1}}(0) \hdots \rho_{-\bm{k_n}}(0)
	\rho_{\bm{k_1}}(t) \hdots \rho_{\bm{k_n}}(t) \rangle}
{\langle \rho_{\bm{-k_1}}(0) \hdots \rho_{-\bm{k_n}}(0)  
	\rho_{\bm{k_1}}(0) \hdots \rho_{\bm{k_n}}(0) \rangle},
\end{equation} 
where $\rho_{\bm{k}}(t)$ is a collective density mode at wavevector $\bm{k}$ and time $t$,
the angle brackets denote an ensemble average, and the label $n$ ($n=1,\hdots,\infty$) specifies the level of the hierarchy.
In the overdamped limit, they satisfy 
\begin{gather} 
\nu_n\dot{\phi}_n(k_1,\hdots,k_n,t) + \Omega^2_n(k_1,\hdots,k_n)\phi_n(k_1,\hdots,k_n,t) 
\nonumber \\ 
+\int_0^t M_n(k_1,\hdots,k_n,u) \dot{\phi}_n(k_1,\hdots,k_n,t-u) du  = 0 \label{eq:GMCTphi_n}, 
\end{gather} 
where $\nu_n$ is an effective friction coefficient, and 
\begin{equation} 
\label{eq:Omega2n}
\Omega^2_n(k_1,\hdots,k_n) = D_0 \left[ \frac{k_1^2}{S(k_1)} + \hdots + \frac{k_n^2}{S(k_n)} \right] 
\end{equation}
are the so-called bare frequencies with $D_0$ denoting the bare diffusion constant, and $S(k)$ are the static structure factors. Note that $\phi_1(k,t)=F(k,t)/S(k)$.
The memory functions are given by
\begin{eqnarray}
\label{eq:Mn} 
M_n(k_1,\hdots,k_n,t) = \frac{\rho D_0}{16\pi^3} \sum_{i=1}^n \frac{\Omega^2_1(k_i)}{\Omega^2_n(k_1,\hdots,k_n)}
\nonumber \\
\times \int d\bm{q} |\tilde{V}_{\bm{q,k}_i-\bm{q}}|^2 
S(q) S(|\bm{k}_i-\bm{q}|)
\hphantom{XXXX}
\nonumber \\
\times \phi_{n+1}(q,|\bm{k}_1-\bm{q}\delta_{i,1}|,\hdots,|\bm{k}_n-\bm{q}\delta_{i,n}|,t) \nonumber \\
\end{eqnarray}
where $\rho$ is the bulk density, $\delta_{i,j}$ is the Kronecker delta
function, and $\tilde{V}_{\bm{q,k}_i-\bm{q}}$ are the static vertices that represent
wavevector-dependent coupling strengths. 
The latter are defined as
\begin{equation}
\label{eq:V}
\tilde{V}_{\bm{q,k-q}} = 
({\hat{\bm{k}}} \cdot \bm{q}) c(q) + 
{\hat{\bm{k}}} \cdot (\bm{k-q}) c(|\bm{k-q}|),
\end{equation}
where $\hat{\bm{k}} = \bm{k}/k$ and $c(q)$ is the direct
correlation function,\cite{hansen1990theory} which is related to the static structure factor as $c(q)
\equiv [1-1/S(q)] / \rho$. 
The initial conditions for Eq.\ (\ref{eq:GMCTphi_n}) are $\phi_n(k_1,\hdots,k_n,0)=1$ for all $n$.

 In order to solve the equations, a closure is necessary for the last included level $N<\infty$. In the absence of a known exact closure, we may approximate the last level $\phi_N$ by the product of $\phi_{N-1}$ and $\phi_1$. To further account for permutation invariance of all wavenumber arguments $\{k_1,\hdots,k_n\}$, we write
\begin{equation}
\phi_N(k_1,\hdots,k_N, t)=\frac{1}{N}\sum_{i=1}^N\phi_1(k_i,t)\times\phi_{N-1}(\{k_j\}^{(N-1)}_{j\neq i},t)
\label{eq:closure_t}
\end{equation}
where $\{k_j\}^{(N-1)}_{j\neq i}$ represents the $N-1$ wavenumbers in
$\{k_1,\hdots,k_N\}$ except the $k_i$. 
This is referred to as a mean-field (MF) closure and is denoted as MF-$N[(N-1)^11^1]$. This closure is qualitatively equivalent to the one used in the accompanying paper\cite{luo2019generalized} but here we explictly link $\phi_N(k,t)$ to $\phi_1(k,t)$ and $\phi_{N-1}(k,t)$ for the convenience of the derivation below. An alternative closure approximation is a simple truncation of the hierarchy such that $\phi_N(k_1,\hdots,k_N,t)=0$, which is equivalent to setting $\phi_{N-1}=\exp(-t/\tau_N)$. We refer to this as an exponential (EXP-$N$) closure. In Ref.\ \onlinecite{janssen2015microscopic} and the accompanying paper\cite{luo2019generalized}, it has been tested numerically that the mean-field and exponential closures provide an upper and lower bound respectively for the relaxation dynamics in the limit of large $N$. As shown in the accompanying paper, the MF-$N$ series manifestly converges faster with $N$ than the EXP-$N$ closure series when close to the glass transition. Hence, we focus solely on the MF closures in the following analysis.

Equations (\ref{eq:GMCTphi_n}), ( \ref{eq:Mn}) and (\ref{eq:closure_t}) define a unique, well-behaved solution for all
$\phi_n(k_1,\hdots,k_n, t)$ with $n\leq N$.\cite{biezemans2020glassy} Although no known analytic result exists for the complete wavevector- and time-dependent dynamics, we can derive several universal properties of the solutions as discussed below. Furthermore, the full solutions may also be found numerically in a self-consistent procedure once the static structure factors (and the corresponding bulk density) of the material of interest are known. Finally, We emphasize that the theory is free from fit parameters, and that no phenomenological assumptions are made regarding the emergence of glassy dynamics.

For convenience we can also rewrite the GMCT equations in complex frequency space using the Laplace transform $F(s)=\mathcal L(f(t))(s)=\int_{0}^{\infty}f(t)e^{-st}dt$, yielding 
\begin{gather}
\frac{s\Phi_n(k_1,\hdots,k_n,s)}{1-s\Phi_n(k_1,\hdots,k_n,s)}=
\frac{s\nu_n+sm_n(k_1,\hdots,k_n,s)}{\Omega_n^2(k_1,\hdots,k_n)}.
\label{eq:GMCT_laplace}
\end{gather}
with the closure MF-$N[(N-1)^11^1]$
\begin{gather} 
s\Phi_N(k_1,\hdots,k_N,s)=
\nonumber \\
\frac{1}{N}\sum_{i=1}^N s\mathcal{L}\big[\phi_1(k_i,t)\times \phi_{N-1}(\{k_j\}^{(N-1)}_{j\neq i},t)\big],
\label{eq:closure_laplace}
\end{gather}
where $\Phi_n(k_1,\hdots,k_n,s)$ and $m_n(k_1,\hdots,k_n,s)$ are the Laplace transformation of $\phi_n(k_1,\hdots,k_n,t)$ and  $M_n(k_1,\hdots,k_n,t)$, respectively. The relation between $m_n(k_1,\hdots,k_n)$ and $\Phi_{n+1}(k_1,\hdots,k_n)$ still satisfies Eq.~(\ref{eq:Mn}).
Since the $\Omega_n(k_1,\hdots,k_n)$ are simply constants in Eq.~(\ref{eq:GMCT_laplace}), we define a new memory function absorbing $\Omega_n(k_1,\hdots,k_n)$, 
\begin{gather} 
\hat{m}_n(k_1,\hdots,k_n,s)=\frac{m_n(k_1,\hdots,k_n,s)}{\Omega_n^2(k_1,\hdots,k_n)}.
\label{eq:hatM}
\end{gather}

By noticing that the memory functions $\hat{m}_n(k_1,\hdots,k_n,s)$ are linear combinations of the next-order density correlators  $\Phi_{n+1}(k_1,\hdots,k_n,s)$ and that
the three-dimensional integration over $d\bm{q}$ in Eq.~(\ref{eq:Mn}) can be transformed to a two-dimensional summation, we obtain the following linear equation
\begin{gather} 
\hat{m}_n(k_1,\hdots,k_n,s)=
\nonumber \\
\sum_{iqp}V_{n+1}(k_1,\hdots ,k_n,q,p,i)\Phi_{n+1}(q,p,\{k_j\}^{(n-1)}_{j\neq i},s),
\label{eq:hatmPhi}
\end{gather}
%
 where $V_{n+1}(k_1,\hdots ,k_n,q,p,i)$ are effective vertex coefficients that depend on the static structure factors and the wavenumber arguments in $\Phi_{n+1}(q,p,\{k_j\}^{(n-1)}_{j\neq i},s)$. The summation $\sum_{iqp}$ represents a double sum over wavenumbers $q$ and $p$ as well as the index $i$ in Eq.~(\ref{eq:Mn}). Explicitly, if we follow the discretization of wavenumbers in Ref.~\onlinecite{franosch1997asymptotic}, the coefficients are
\begin{gather}
\label{eq:Vcoeff}
V_{n+1}(k_1,\hdots ,k_n,q,p,i)=\frac{\rho D_0 h^5}{32\pi^2} \frac{\Omega^2_1(k_i)}{\Omega^4_n(k_1,\hdots,k_n)}
\nonumber \\
\frac{\hat{q}\hat{p}}{\hat{k}_i^3}\left[(\hat{q}^2-\hat{p}^2+\hat{k}_i^2)c(q)+(\hat{p}^2-\hat{q}^2+\hat{k}_i^2)c(p)\right]^2 S(q)S(p)
\nonumber \\
\end{gather}
and 
$$\sum_{iqp}=\sum_{i=1}^n \sum_{\hat{q}=1/2}^{M-1/2} \sum_{\hat{p}=|\hat{k}_i-\hat{q}|+1/2}^{\hat{k}_i+\hat{q}-1/2}$$ 
where $q,p,k_i$ are integer indices $1,2,\hdots,M$ and $\hat{q},\hat{p},\hat{k_i}$ are the corresponding half integers $1/2,3/2,\hdots,M-1/2$. $M$ is the total number of wavenumbers and $h$ is the step size between all equally spaced wavenumbers.
In the following, we will discuss the asymptotic solutions of the GMCT equations based on Eqs.~(\ref{eq:GMCT_laplace}), (\ref{eq:closure_laplace}) and (\ref{eq:hatmPhi}).

\section{Long-time limit}
\label{sec:long}
We first present the derivation of the asymptotic solutions for the long-time limits of the $2n$-point density correlation functions, in which we introduce the technique of asymptotic expansions also used in the time-dependent GMCT solutions.
Applying the final value theorem $\lim_{t\to\infty}f(t)=\lim_{s\to0}sF(s)$ to Eq.~(\ref{eq:GMCT_laplace}), the long-time limits of the correlators $f_n(k_1,\hdots,k_n)\equiv \lim_{t\to\infty}\phi_n(k_1,\hdots,k_n,t)$ satisfy
\begin{gather}
\label{eq:GMCT_long}
\frac{f_n(k_1,\hdots,k_n)}{1-f_n(k_1,\hdots,k_n)}=\mathcal{M}_n(k_1,\hdots,k_n),
\end{gather}
where
\begin{gather}
\mathcal{M}_n(k_1,\hdots,k_n) = 
\nonumber \\
\sum_{iqp}V_{n+1}(k_1,\hdots ,k_n,q,p,i)f_{n+1}(q,p,\{k_j\}^{(n-1)}_{j\neq i}). 
\label{eq:Mn_longtime} 
\end{gather}
The closure Eq.~(\ref{eq:closure_laplace}) becomes 
\begin{equation}
\label{eq:closure_long}
f_N(k_1,\hdots,k_N)=\frac{1}{N}\sum_{i=1}^N f_1(k_i)\times f_{N-1}(\{k_j\}^{(N-1)}_{j\neq i}).
\end{equation}
Given the structure factors $S(k)$, we can calculate the form factors
$f_n(k_1,\hdots,k_n)$ $(1 \leq n \leq N)$ iteratively.
There is a critical packing fraction $\varphi^c$ separating the liquid from the glass solutions under the closure MF-$N[(N-1)^11^1]$, i.e., the point $\varphi=\varphi^c$ is a glass-transition singularity which marks a bifurcation point.
For negative values of the reduced packing fraction $\epsilon=(\varphi-\varphi^c)/\varphi^c$, all $f_n(k_1,\hdots,k_n)$ are zero, while for $\epsilon\geq 0$ ideal glass states with $f_n(k_1,\hdots,k_n)>0$ are obtained. For $\varphi$ approaching the critical value from above, the $f_n(k_1,\hdots,k_n)$ approach positive constants, called critical form factors $f_n^c(k_1,\hdots,k_n)$. Note that when $n=1$, $f_1^c(k_1)$ are the so-called non-ergodicity parameters in standard MCT. 
We assume that the bifurcation in $f_n(k_1,\hdots,k_n)$ at the critical point is of the type $A_2$, which is the common case in standard MCT.\cite{gotze2008complex,arnol2003catastrophe} The PY hard-sphere system indeed conforms to this bifurcation scenario, as shown in the accompanying paper\cite{luo2019generalized} and in the numerical analysis below. We point out that other possible singularities of type $A_l$ with $l>2$ do exist in MCT\cite{dawson2000higher,gotze1991liquids} and are also expected to exist in GMCT on mathematical grounds; our preliminary GMCT work on multicomponent systems numerically confirms this.

\begin{figure}
	\epsfig{file=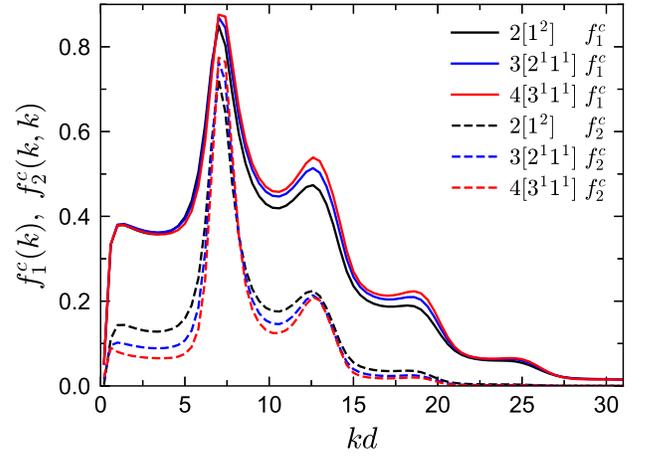,width=0.46\textwidth}
	\caption{\label{fig:fc} 
		The critical form factors $f_1^c(k)$ and $f_2^c(k,k)$ as a function of wavenumber $k$ at the critical packing fraction 
                $\varphi^c$ for different GMCT MF closure levels. 
                Solid lines are the non-ergodicity parameters $f_1(k)$ at critical packing fractions $0.515914$, $0.531888$ and
 $0.544172$ for MF-$N[(N-1)^11^1]$ closure levels with $N=2,3,4$, respectively.
                Dashed lines are $f_2^c(k,k)$ at the same corresponding critical packing fractions.               
	}
\end{figure}

Figure \ref{fig:fc} shows the critical form factors for PY hard spheres under closures MF-$N[(N-1)^11^1]$ with $N=2,3,4$. We find that both $f^c_1(k)$ and $f^c_2(k,k)$ are modulated by the structure of $S(k)$ with a maximum at $kd\approx7.4$, where $d$ denotes the hard-sphere diameter. As discussed in the accompanying paper,\cite{luo2019generalized} increasing the closure level $N$ leads to overall higher non-ergodicity parameters $f^c_1(k)$ (solid lines in Fig.~\ref{fig:fc}) which physically corresponds to relatively slower relaxation dynamics. However, $f^c_2(k,k)$ shows overall the opposite trend (dashed lines in Fig.~\ref{fig:fc}), i.e.\ a higher closure level $N$ leads to a lower $f_2^c(k,k)$, except at the wavenumbers around the first peak of $S(k)$. Hence $[f^c_1(k)]^2\neq f_2^c(k,k)$ for closure levels $N>2$, a result that is consistent with the time-dependent GMCT results for weakly polydisperse hard spheres.\cite{janssen2015microscopic} We may interpret this inequality as an indication for dynamical heterogeneity, since $f_2^c(k,k)$ is akin to a variance of the density fluctuations contained in $f^c_1(k)$. We point out, however, that more studies are needed to accurately link our theory to the heterogeneity properties such as the four-point susceptibility $\chi_4(k,t)$. 

In order to study the form factors $f_n(k_1,\hdots,k_n)$ near the critical point when $\varphi>\varphi^c$, we write 
\begin{gather}
f_n(k_1,\hdots,k_n)=f_n^c(k_1,\hdots,k_n)
\nonumber \\
+\left[1-f_n^c(k_1,\hdots,k_n)\right]^2r_n(k_1,\hdots,k_n)
\end{gather} 
and solve Eqs.~(\ref{eq:GMCT_long})-(\ref{eq:closure_long}) for small $r_n(k_1,\hdots,k_n)$ and small positive $\epsilon$. From Eq.~(\ref{eq:GMCT_long}) and (\ref{eq:Mn_longtime}) we obtain that for $n<N$,
\begin{gather}
r_n(k_1,\hdots,k_n)
\nonumber \\
-\sum_{iqp}
C_{n+1}^c(k_1,\hdots ,k_n,q,p,i)r_{n+1} (q,p,\{k_j\}^{(n-1)}_{j\neq i})
\nonumber \\
=\bigg\{\sum_{iqp}V_{n+1}^{'c}(k_1,\hdots ,k_n,q,p,i)f^c_{n+1}(q,p,\{k_j\}^{(n-1)}_{j\neq i})\epsilon 
\nonumber \\
-\left[1-f^c_n(k_1,\hdots,k_n)\right]r^2_n(k_1,\hdots,k_n)\bigg\}
\nonumber \\
+\bigg\{\sum_{iqp}C_{n+1}^{'c}(k_1,\hdots ,k_n,q,p,i)\epsilon r_{n+1} (q,p,\{k_j\}^{(n-1)}_{j\neq i})
\nonumber \\
- \left[1-f^c_n(k_1,\hdots,k_n)\right]^2r^3_n(k_1,\hdots,k_n)\bigg\},
\nonumber \\
+O(\epsilon^2,r^4,\epsilon r^2)
\label{eq:g_first}
\end{gather}
where
\begin{eqnarray} 
&&V_{n+1}^{'c}(k_1,\hdots ,k_n,q,p,i)=\left.\frac{\partial V_{n+1}(k_1,\hdots ,k_n,q,p,i)}{\partial \epsilon}\right\vert_{\varphi=\varphi^c},
\nonumber \\
\end{eqnarray}
\begin{eqnarray} 
&&C_{n+1}^c(k_1,\hdots ,k_n,q,p,i)=
\nonumber \\
&&V_{n+1}^c(k_1,\hdots ,k_n,q,p,i)\left[1-f^c_{n+1}(q,p,\{k_j\}^{(n-1)}_{j\neq i})\right]^2,
\nonumber \\
\end{eqnarray}
and 
\begin{eqnarray} 
&&C_{n+1}^{'c}(k_1,\hdots ,k_n,q,p,i)=
\nonumber \\
&&V_{n+1}^{'c}(k_1,\hdots ,k_n,q,p,i)\left[1-f^c_{n+1}(q,p,\{k_j\}^{(n-1)}_{j\neq i})\right]^2.
\nonumber \\
\end{eqnarray}
The first crucial assumption in our derivation is that all $V_{n+1}(k_1,\hdots,k_n,q,p,i)$ [and $C_{n+1}(k_1,\hdots,k_n,q,p,i)$] vary smoothly with $\epsilon$, hence we have applied $V_{n+1}(k_1,\hdots,k_n,q,p,i)\approx V^{c}_{n+1}(k_1,\hdots,k_n,q,p,i)+V^{'c}_{n+1}(k_1,\hdots,k_n,q,p,i)\epsilon$ in the above equations. This assumption is reasonable for PY hard spheres since near the critical point, the static structure factor can be fairly accurately described by $S(k)\approx S^c(k)+\epsilon S^{(1)}(k)$, where $S^{(1)}(k)$ is a constant for a given critical packing fraction $\varphi^c$.
From the closure Eq.~(\ref{eq:closure_long}), we obtain at level $N$
\begin{eqnarray} 
&&\left[1-f^c_N(k_1,\hdots,k_N)\right]^2r_N(k_1,\hdots,k_N)=\frac{1}{N}\sum_{i=1}^N
\nonumber \\
&&\bigg\{\big[\left[1-f^c_1(k_i)\right]^2f^c_{N-1}(\{k_j\}^{(N-1)}_{j\neq i})r_1(k_i)
\nonumber \\
&&+\left[1-f^c_{N-1}(\{k_j\}^{(N-1)}_{j\neq i})\right]^2f^c_1(k_i)r_{N-1}(\{k_j\}^{(N-1)}_{j\neq i})\big]
\nonumber \\
&&+\left[1-f^c_1(k_i)\right]^2\left[1-f^c_{N-1}(\{k_j\}^{(N-1)}_{j\neq i})\right]^2
\nonumber \\
&&\times r_1(k_i)r_{N-1}(\{k_j\}^{(N-1)}_{j\neq i})\bigg\}.
\nonumber \\
\label{eq:g_last}
\end{eqnarray}
Expanding $r_n(k_1,\hdots,k_n)= A_n(k_1,\hdots,k_n)\sqrt{\epsilon}+B_n(k_1,\hdots,k_n)\epsilon+ D_n(k_1,\hdots,k_n)\epsilon^{3/2}$ as an ansatz, we can estimate the coefficients $A_n(k_1,\hdots,k_n)$ and $B_n(k_1,\hdots,k_n)$ by solving Eq.~(\ref{eq:g_first}) and (\ref{eq:g_last}).

 To order $\sqrt\epsilon$, we obtain $N$ linear equations 
\begin{eqnarray} 
&&A_n(k_1,\hdots,k_n)
\nonumber \\
&&=\sum_{iqp}C_{n+1}^c(k_1,\hdots ,k_n,q,p,i)A_{n+1} (q,p,\{k_j\}^{(n-1)}_{j\neq i}),
\nonumber \\
&&\hspace{55mm}\text{for} \ \ \ n<N
\label{eq:g_first_linear}
\end{eqnarray}
and 
\begin{eqnarray} 
&&A_N(k_1,\hdots,k_N)=\frac{1}{N\left[1-f^c_N(k_1,\hdots,k_N)\right]^2}\sum_{i=1}^N
\nonumber \\
&&\bigg\{\left[1-f^c_1(k_i)\right]^2f^c_{N-1}(\{k_j\}^{(N-1)}_{j\neq i})A_1(k_i)
\nonumber \\
&&+\left[1-f^c_{N-1}(\{k_j\}^{(N-1)}_{j\neq i})\right]^2f^c_1(k_i)A_{N-1}(\{k_j\}^{(N-1)}_{j\neq i})\bigg\}
\nonumber \\
\label{eq:g_last_linear}
\end{eqnarray}
If we rewrite all $A_n$ as a vector 
\begin{eqnarray*}
\bm{A}=\bigg[&&A_1(1),A_1(2),\hdots, A_1(M),\\ &&A_2(1,1),A_2(1,2),\hdots,A_2(M,M),\\
&&\hdots,\\
&&A_N(1,\hdots,1),A_N(1,\hdots,2),\hdots,A_N(M,\hdots,M)\bigg]^T,
\end{eqnarray*}
with $M+M^2+\hdots+M^N$ elements,
and construct a $(M+M^2+\hdots+M^N)\times(M+M^2+\hdots+M^N)$ matrix $\bm{C}$ which contains all the coefficients of the corresponding $\bm{A}$ terms, then Eq.~(\ref{eq:g_first_linear}) and (\ref{eq:g_last_linear}) are equivalent to an equation to calculate the eigenvector of the matrix $\bm{C}$,
\begin{equation}
\bm{CA}=\bm{A}.
\label{eq:CAA}
\end{equation}
Note that the coefficient matrix $\bm{C}$ only depends on the information of the critical point. The Jacobian matrix of the system of Eqs.~(\ref{eq:GMCT_long}), (\ref{eq:Mn_longtime}) and (\ref{eq:closure_long}) at the critical point is equivalent to the matrix $\bm{1}-\bm{C}$. It can be seen from Eqs.~(\ref{eq:g_first_linear}) and (\ref{eq:g_last_linear}) that all elements of $\bm{C}$ are non-negative. Thus there is a nondegenerate maximum eigenvalue $E$ of the matrix $\bm{C}$ according to the Frobenius-Perron theorem.\cite{meyer2000matrix} For the glass-transition singularity at the critical point we know that $E=1$, which is obvious from Eq.~(\ref{eq:CAA}).
We denote the right eigenvector and left eigenvector corresponding to the eigenvalue 1 of $\bm{C}$ as $\bm{e}$ and $\hat{\bm{e}}^T$, respectively, i.e.
\begin{equation}
\bm{Ce}=\bm{e}; \ \ \hat{\bm{e}}^T\bm{C}=\hat{\bm{e}}^T. 
\label{eq:left_vector}
\end{equation}
To fix the eigenvectors uniquely we impose the convention
\begin{equation}
\hat{\bm{e}}^T\bm{e}=1
\label{eq:con1}
\end{equation} and 
\begin{eqnarray} 
\sideset{}{'}\sum_{n<N}\hat{e}_n(k_1,\hdots,k_n)
\left[1-f^c_n(k_1,\hdots,k_n)\right]e^{2}_n(k_1,\hdots,k_n)=1,
\nonumber\\
\label{eq:con2}
\end{eqnarray}
where
$\sideset{}{'}\sum_{n=m}$ represents the summation over all possible wavenumbers ${k_1,\hdots,k_m}$ for the level $m$.
Therefore, the coefficients of order $\sqrt{\epsilon}$ are 
\begin{equation}
A_n(k_1,\hdots,k_n)=Ae_n(k_1,\hdots,k_n)
\end{equation} 
where $A$ is an overall factor to be determined. 

In order to estimate $A$, the next order $\epsilon$ of Eqs.~(\ref{eq:g_first}) and (\ref{eq:g_last}) needs to be considered, which yields
\begin{eqnarray} 
&&B_n(k_1,\hdots,k_n)-
\nonumber \\
&&\sum_{iqp}
C_{n+1}^c(k_1,\hdots ,k_n,q,p,i)B_{n+1} (q,p,\{k_j\}^{(n-1)}_{j\neq i})
\nonumber \\
&&=\sum_{iqp}V_{n+1}^{'c}(k_1,\hdots ,k_n,q,p,i)f^c_{n+1}(q,p,\{k_j\}^{(n-1)}_{j\neq i})
\nonumber \\
&&-\left[1-f^c_n(k_1,\hdots,k_n)\right]A^2_n(k_1,\hdots,k_n)
\nonumber \\
&&\hspace{55mm}\text{for} \ \ \ n<N
\label{eq:g_first_B}
\end{eqnarray}
and for the closure level $N$
\begin{eqnarray} 
&&B_N(k_1,\hdots,k_N)-\frac{1}{N\left[1-f^c_N(k_1,\hdots,k_N)\right]^2}\sum_{i=1}^N
\nonumber \\
&&\bigg\{\left[1-f^c_1(k_i)\right]^2f^c_{N-1}(\{k_j\}^{(N-1)}_{j\neq i})B_1(k_i)
\nonumber \\
&&+\left[1-f^c_{N-1}(\{k_j\}^{(N-1)}_{j\neq i})\right]^2f^c_1(k_i)B_{N-1}(\{k_j\}^{(N-1)}_{j\neq i})\bigg\}
\nonumber \\
&&=\frac{1}{N\left[1-f^c_N(k_1,\hdots,k_N)\right]^2}\sum_{i=1}^N\bigg\{A_1(k_i)A_{N-1}(\{k_j\}^{(N-1)}_{j\neq i})
\nonumber \\
&&\times\left[1-f^c_1(k_i)\right]^2\left[1-f^c_{N-1}(\{k_j\}^{(N-1)}_{j\neq i})\right]^2\bigg\}.
\nonumber \\
\label{eq:g_last_B}
\end{eqnarray}
Written in matrix form, the coefficients of the vector $\bm{B}$ are also the matrix $\bm{C}$ and the remaining terms are the right-hand side of Eqs.~(\ref{eq:g_first_B}) and (\ref{eq:g_last_B}) denoted as a vector $\bm{R}$, thus
\begin{equation}
\bm{(1-C)B}=\bm{R}.
\label{eq:BR}
\end{equation}
 Notice that the second equation in Eq.~(\ref{eq:left_vector}) indicates that
 \begin{equation}
 \bm{\hat{e}}^T\bm{(1-C)X}=0
 \label{eq:ehat}
 \end{equation} for any column vector $\bm{X}$. Thus the factor $A$ can be determined via $\bm{\hat{e}}^T\bm{R}=0$,
\begin{equation} 
A=\sqrt{ \frac{\sigma }{1-\lambda} } 
\label{eq:a}
\end{equation}
where 
\begin{eqnarray} 
&&\sigma=\sideset{}{'}\sum_{n<N}\hat{e}_n(k_1,\hdots,k_n)
\nonumber \\
&& \times \sum_{iqp}V_{n+1}^{'c}(k_1,\hdots ,k_n,q,p,i)f^c_{n+1}(q,p,\{k_j\}^{(n-1)}_{j\neq i})
\label{eq:sigma}
\end{eqnarray}
and
\begin{gather} 
\lambda=\sideset{}{'}\sum_{N}\hat{e}_N(k_1,\hdots,k_N)\frac{1}{\left[1-f^c_N(k_1,\hdots,k_N)\right]^2}\frac{1}{N}
\nonumber \\
\sum_{i=1}^N\left[1-f^c_1(k_i)\right]^2\left[1-f^c_{N-1}(\{k_j\}^{(N-1)}_{j\neq i})\right]^2
\nonumber \\
\times e_1(k_i)e_{N-1}(\{k_j\}^{(N-1)}_{j\neq i}).
\nonumber \\
\label{eq:lambda}
\end{gather}
We restrict our discussion to so-called $A_2$ singularities so that $\lambda<1$.
In this way we obtain the leading asymptotic expression for the 
long-time limits of the density correlators,
\begin{equation}
f_n(k_1,\hdots,k_n)=f_n^c(k_1,\hdots,k_n)+h_n(k_1,\hdots,k_n)\sqrt{\frac{\sigma \epsilon}{1-\lambda}}
\label{eq:f1order}
\end{equation} 
where the critical amplitudes are given by
\begin{equation}
h_n(k_1,\hdots,k_n)=\left[1-f_n^c(k_1,\hdots,k_n)\right]^2e_n(k_1,\hdots,k_n).
\label{eq:hn}
\end{equation}

\begin{figure}
	\epsfig{file=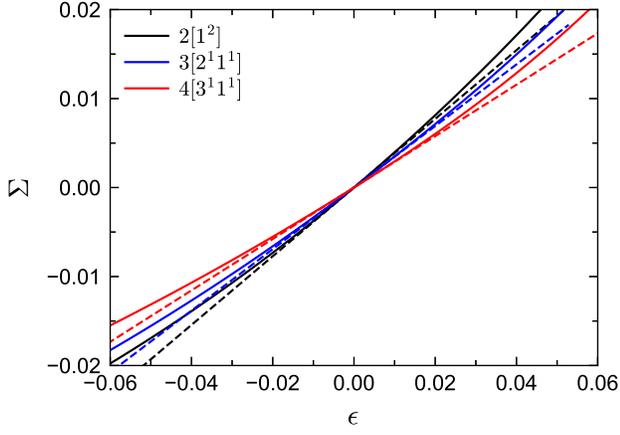,width=0.46\textwidth}
	\caption{\label{fig:Sigma} 
		The separation $\Sigma \approx \sigma \epsilon$ as a function of $\epsilon$. 
                Solid lines are the numerical separation $\Sigma$ for MF-$N[(N-1)^11^1]$ closure levels with $N=2,3,4$, respectively.
                Dashed lines are $\sigma\epsilon$ with the corresponding $\sigma$ in Table \ref{tab:coeff}.               
	}
\end{figure}
\begin{figure}
	\epsfig{file=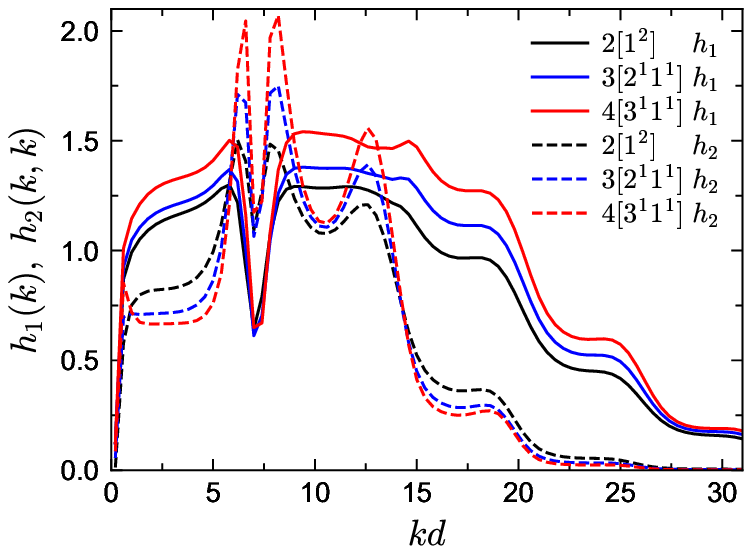,width=0.46\textwidth}
	\caption{\label{fig:h} 
		The critical amplitudes $h_1(k)$ and $h_2(k,k)$ as a function of wavenumber $k$ under different GMCT MF closure levels. 
                Solid lines are $h_1(k)$ and dashed lines are $h_2(k,k)$ under MF-$N[(N-1)^11^1]$ closure levels with $N=2,3,4$.              
	}
\end{figure}

In order to obtain the $\sigma$ for PY hard spheres, we numerically calculate the separation
\begin{eqnarray} 
&&\Sigma=\sideset{}{'}\sum_{n<N}\big[\hat{e}_n(k_1,\hdots,k_n)
\nonumber \\
&& \times \sum_{iqp}V_{n+1}(k_1,\hdots ,k_n,q,p,i)f^c_{n+1}(q,p,\{k_j\}^{(n-1)}_{j\neq i})\big]
\nonumber \\
&&- \sideset{}{'}\sum_{n<N}\big[\hat{e}_n(k_1,\hdots,k_n)
\nonumber \\
&& \times \sum_{iqp}V_{n+1}^{c}(k_1,\hdots ,k_n,q,p,i)f^c_{n+1}(q,p,\{k_j\}^{(n-1)}_{j\neq i})\big]
\nonumber \\
\label{eq:Sigma}
\end{eqnarray}
as a function of $\epsilon$. It can be seen from Fig.~\ref{fig:Sigma} that $\Sigma\approx\sigma\epsilon$ for all MF closures, which agrees with our assumption that the $V_n$ vary smoothly with $\epsilon$. This property allows us to use $\epsilon$ as an order parameter.
Note that the notation $\sigma$ in Ref.\ \onlinecite{franosch1997asymptotic} corresponds to the $\Sigma$ here, which is proportional to $\epsilon$, while the $\sigma$ we use here is a constant. The values of $\sigma$ and $\lambda$ for PY hard spheres under different GMCT closures are listed in Table \ref{tab:coeff}.

In Eq.~(\ref{eq:BR}), $\bm{R}$ only depends on $\bm{A}$. Hence once we know $A_n(k_1,\hdots,k_n)$, a special solution for $\bm{B}$ can be calculated via $\bm{B_0}=(\bm{1}-\bm{C})^{-1}\bm{R}$. 
General solutions of $\bm{B}$ are the linear combinations of the special solution $\bm{B_0}$ and $\bm{A}$, i.e. $\bm{B}=\bm{B_0}+\kappa\bm{A}$. To determine $\kappa$ the order $\epsilon^{3/2}$ of Eqs.~(\ref{eq:g_first}) and (\ref{eq:g_last}) has to be considered. Again, by virtue of Eq.~(\ref{eq:ehat}) and the conventions of $\bm{e}$ and $\bm{\hat e}$ in Eqs.~(\ref{eq:con1}) and (\ref{eq:con2}),  we obtain 
\begin{gather}
\kappa=\frac{1}{2 \sqrt{\sigma(1-\lambda)}}\bigg\{ \sideset{}{'}\sum_{n<N}\hat{e}_n(k_1,\hdots,k_n)\big[
\nonumber \\
-2\left[1-f^c_n(k_1,\hdots,k_n)\right]e_n(k_1,\hdots,k_n)B_{0n}(k_1,\hdots,k_n)
\nonumber \\
+\sum_{iqp}C_{n+1}^{'c}(k_1,\hdots ,k_n,q,p,i)
e_{n+1} (q,p,\{k_j\}^{(n-1)}_{j\neq i})
\nonumber \\
-\frac{\sigma}{1-\lambda}\left[1-f^c_n(k_1,\hdots,k_n)\right]^2e_n^3(k_1,\hdots,k_n)\big]
\nonumber \\
+\sideset{}{'}\sum_{N}\hat{e}_N(k_1,\hdots,k_N)\frac{1}{\left[1-f^c_N(k_1,\hdots,k_N)\right]^2}\frac{1}{N}\times \sum_{i=1}^N
\nonumber\\
\times\left[1-f^c_1(k_i)\right]^2\left[1-f^c_{N-1}(\{k_j\}^{(N-1)}_{j\neq i})\right]^2
\nonumber\\
\times[e_1(k_i)B_{0N-1}(\{k_j\}^{(N-1)}_{j\neq i})
\nonumber\\
+e_{N-1}(\{k_j\}^{(N-1)}_{j\neq i})B_{01}(k_i)]\bigg\}.
\nonumber \\
\label{eq:kappa}
\end{gather}
\begin{figure}
	\epsfig{file=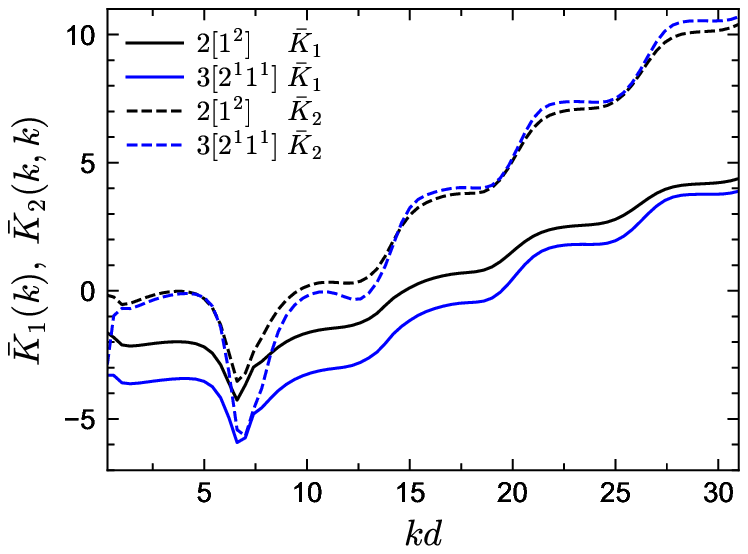,width=0.46\textwidth}
	\caption{\label{fig:Knbar} 
		The amplitudes $\bar{K}_1(k)$ and $\bar{K}_2(k,k)$ as a function of wavenumber $k$ under different GMCT MF closure levels. 
                Solid lines are $\bar{K}_1(k)$ and dashed lines are $\bar{K}_2(k,k)$ for MF-$N[(N-1)^11^1]$ closure levels with $N=2,3$.              
	}
\end{figure}
Therefore, the leading preasymptotic corrections for the form factors are
\begin{eqnarray}
&&f_n(k_1,\hdots,k_n)=f_n^c(k_1,\hdots,k_n)
\nonumber\\
&&+h_n(k_1,\hdots,k_n)\sqrt{\frac{\sigma }{1-\lambda}}\sqrt\epsilon
\big[1+ (\bar{K}_n(k_1,\hdots,k_n)+\kappa)\sqrt\epsilon\big],
\nonumber\\
\label{eq:long_time_asy}
\end{eqnarray}
where 
\begin{equation}
\label{eq:Kbar}
\bar{K}_n(k_1,\hdots,k_n)=\sqrt{\frac{1-\lambda}{\sigma}}\frac{B_{0n}(k_1,\hdots,k_n)}{e_n(k_1,\hdots,k_n)}.
\end{equation}
These results for the long-time limit of multi-point density correlators are similar to the MCT results, but all parameters are different and depend on the MF closures we apply. It is clear now that only when $\epsilon\ll{\left[\bar{K}_n(k_1,\hdots,k_n)+\kappa\right]^{-2}}$ the leading results are applicable.

We show the amplitudes $h_1(k)$, $h_2(k,k)$ in Fig.~\ref{fig:h} and $\bar{K}_1(k)$, $\bar{K}_2(k,k)$ in Fig.~\ref{fig:Knbar} under different closures for PY hard spheres. Because of the large size of the matrix $\bm{C}$,  $\bar{K}_n(k_1,\hdots,k_n)$ is only calculated up to $N=3$ within reasonable computing time, but $h_n(k_1,\hdots,k_n)$ is successfully calculated up to $N=4$ since we can iteratively calculate the largest real eigenvalue and the corresponding eigenvectors of $\bm{C}$ relatively fast. It can be seen that $h_1(k)$, $h_2(k,k)$, $\bar{K}_1(k)$, and $\bar{K}_2(k,k)$ are all modulated by the structure of $S(k)$, similar to $f^c_1(k)$ and $f^c_2(k,k)$ in Fig.~\ref{fig:fc}. Moreover, the trend of $h_1$ and $\bar{K}_1$ over closure level $N$ is regular, in the sense that $h_1(k)$ [$\bar{K}_1(k)$] increases (decreases) simultaneously for all wavenumbers at least up to $kd=30$. However, $h_2(k,k)$ and $\bar{K}_2(k,k)$ are more complex and there are crossovers under different closure levels (dashed lines in Fig.~\ref{fig:h} and dotted lines in Fig.~\ref{fig:Knbar} at $kd\approx15$). This may also be a representation of  dynamical heterogeneity, as a consequence of the higher-order density correlators included in our theory. We point out that with the increase of wavenumber $k$ when $kd>15$, $\bar{K}_2(k,k)+\kappa$ considerably grows, which indicates that the applicability range of the leading asymptotic solution Eq.~(\ref{eq:f1order}) at higher wavenumbers is much narrower. 

\begin{figure}
	\epsfig{file=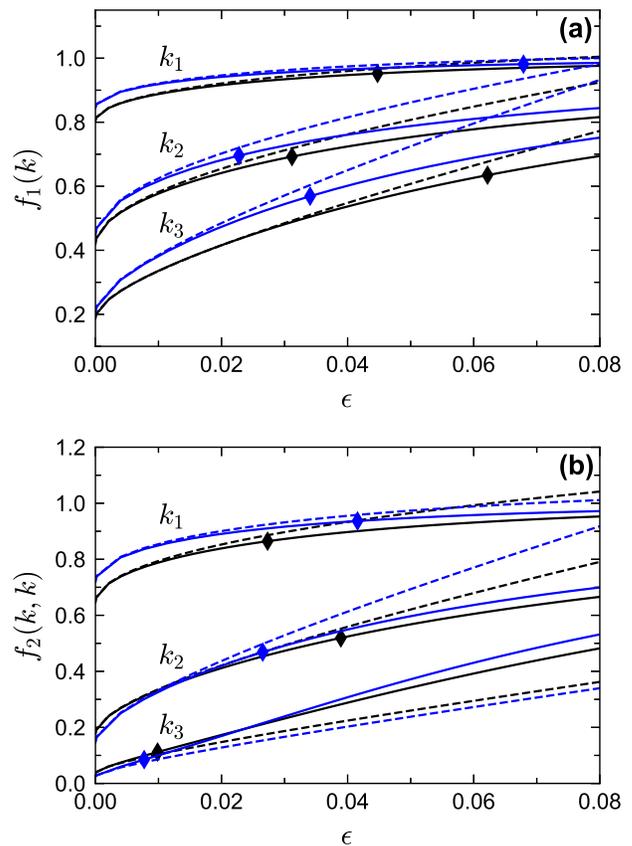,width=0.46\textwidth}
	\caption{\label{fig:f} 
		The form factors for the wavenumbers $k_1d=7.4$, $k_2d=10.6$, and $k_3d=17.4$ under MF-$N[(N-1)^11^1]$ closures with $N=2$ (black lines) and $N=3$ (blue lines). 
                \textbf{(a)} The form factors $f_1(k)$. Solid lines are the numerical solutions and dashed lines are the leading and next-to-leading asymptotic results using Eq.~(\ref{eq:long_time_asy}). Diamonds marks the point with relative error $10\%$ for $f_1(k)-f_1^c(k)$.   \textbf{(b)} The form factors $f_2(k,k)$.   Diamonds marks the point with relative error $10\%$ for $f_2(k,k)-f_2^c(k,k)$.        
	}
\end{figure}

To test the asymptotic solutions with leading preasymptotic corrections, we show the numerical solutions of $f_1(k)$, $f_2(k,k)$ (solid lines in Fig.~\ref{fig:f}) and the corresponding predicted results from Eq.~(\ref{eq:long_time_asy}) (dashed lines in Fig.~\ref{fig:f}). We check three wavenumbers $k_1d=7.4$, $k_2d=10.6$, and $k_3d=17.4$. At all wavenumbers, Eq.~(\ref{eq:long_time_asy}) works well up to at least $\epsilon\approx0.01$ [the lines labeled by $k_3$ in Fig.~\ref{fig:f}(b)] and even up to $\epsilon\approx0.06$ in some cases [blue line labeled by $k_1$ in Fig.~\ref{fig:f}(a)]; the validity range is also indicated by the diamond symbols in Fig.~\ref{fig:f}, which mark a $10\%$ relative error for $f-f^c$.
We also find that for a given wavenumber $k$, the applicability range of the $\epsilon$ expansion does not change much for a higher closure level, suggesting that our results are valid for arbitrary MF closure levels. 
Overall, this establishes Eq.~(\ref{eq:long_time_asy}) as the general asymptotic result with leading corrections for the long-time limit of multi-point density correlation functions.
\section{Dynamics at the critical point}
Now let us discuss the dynamics at the critical point. Here the critical point means $\epsilon\rightarrow0^-$, i.e.\ we approach the transition from the liquid side in order to capture the full two-step relaxation. Note that for $\epsilon\rightarrow0^+$ the first relaxation step, i.e.\ the critical decay toward the $\beta$ regime, is also the same as for $\epsilon\rightarrow0^-$.  
We define a unique time scale $\tau_\beta$ which separates the two decays and which also characterizes the $\beta$-relaxation regime, via $\phi_1(k,\tau_\beta)=\phi_1^c(k)$ for any given $k$. 
As tested in the accompanying paper,\cite{luo2019generalized} there are two exponents $a$ and $b$ characterizing the power-laws of the critical decay and the von Schweidler law in the early and late $\beta$-relaxation regime, respectively. These two exponents obey the non-trivial relation $\lambda=\Gamma(1-a)^2/\Gamma(1-2a)=\Gamma(1+b)^2/\Gamma(1+2b)$. In the following we will verify that this relation is rigorously preserved within GMCT under MF closures. 

The GMCT equations (\ref{eq:GMCT_laplace})
can be simplified for the slow dynamics near the critical point, since the $\nu_n$ in Eq.~(\ref{eq:GMCT_laplace}) can be ignored compared to $m_n$. Now Eq.~(\ref{eq:GMCT_laplace}) and (\ref{eq:hatmPhi}) become
\begin{gather}
\frac{s\Phi_n(k_1,\hdots,k_n,s)}{1-s\Phi_n(k_1,\hdots,k_n,s)}=
s\hat{m}_n(k_1,\hdots,k_n,s)
\nonumber \\
=\sum_{iqp}V_{n+1}(k_1,\hdots ,k_n,q,p,i)s\Phi_{n+1}(q,p,\{k_j\}^{(n-1)}_{j\neq i},s) 
\label{eq:GMCT_laplace_3}
\end{gather}  
Notice that Eq.~(\ref{eq:GMCT_laplace_3}) is time-scale invariant and therefore this equation alone cannot define a unique time scale.

Similar to the long-time case, we first introduce a function $g_n(k_1,\hdots,k_n,t)$ and its Laplace transform $G_n(k_1,\hdots,k_n,s)$ such that the correlators are represented as 
\begin{eqnarray} 
\phi_n(k_1,\hdots ,k_n,t)-f_n^c(k_1,\hdots ,k_n)=
\nonumber \\
\left[1-f_n^c(k_1,\hdots ,k_n)\right]^2g_n(k_1,\hdots ,k_n,t)
\label{eq:approx_gt}
\end{eqnarray}
in the time domain or
\begin{eqnarray} 
s\Phi_n(k_1,\hdots ,k_n,s)-f_n^c(k_1,\hdots ,k_n)=
\nonumber \\
\left[1-f_n^c(k_1,\hdots ,k_n)\right]^2sG_n(k_1,\hdots ,k_n,s)
\label{eq:approx_gs}
\end{eqnarray}
in the frequency domain.
The functions $g_n(k_1,\hdots ,k_n,t)$ and $sG_n(k_1,\hdots ,k_n,s)$ are the generalizations of the time-independent values $r_n(k_1,\hdots ,k_n)$ that appeared in the previous section. Both functions reduce to $r_n(k_1,\hdots ,k_n)$ in the long-time limit for $\epsilon>0$.
Substituting Eq.~(\ref{eq:approx_gs}) into Eq.~(\ref{eq:GMCT_laplace_3})
we obtain similar results to Eq.~(\ref{eq:g_first}), only replacing $f_n(k_1,\hdots,k_n)$ by $s\Phi_n(k_1,\hdots,k_n,s)$  and replacing $r_n(k_1,\hdots,k_n)$ by $sG_n(k_1,\hdots,k_n,s)$, which read
\begin{eqnarray} 
&&sG_n(k_1,\hdots,k_n,s)-
\nonumber \\
&&\sum_{iqp}
C_{n+1}^c(k_1,\hdots ,k_n,q,p,i)sG_{n+1} (q,p,\{k_j\}^{(n-1)}_{j\neq i},s)
\nonumber \\
&&=\bigg\{\sum_{iqp}V_{n+1}^{'c}(k_1,\hdots ,k_n,q,p,i)f^c_{n+1}(q,p,\{k_j\}^{(n-1)}_{j\neq i})\epsilon 
\nonumber \\
&&-\left[1-f^c_n(k_1,\hdots,k_n)\right]s^2G^2_n(k_1,\hdots,k_n,s)\bigg\}
\nonumber \\
&&+\bigg\{\sum_{iqp}C_{n+1}^{'c}(k_1,\hdots ,k_n,q,p,i)\epsilon sG_{n+1} (q,p,\{k_j\}^{(n-1)}_{j\neq i},s)
\nonumber \\
&&- \left[1-f^c_n(k_1,\hdots,k_n)\right]^2s^3G^3_n(k_1,\hdots,k_n,s)\bigg\} 
\nonumber \\
&&+O(\epsilon^2,(sG)^4,\epsilon(SG)^2)
\nonumber\\
&&\hspace{55mm}\text{for} \ \ \ n<N
.
\label{eq:G_first}
\end{eqnarray}
However, the closure Eq.~(\ref{eq:closure_laplace}) requires more careful consideration and becomes 
\begin{eqnarray} 
\label{eq:G_last}
&&sG_N(k_1,\hdots,k_N,s)
-\frac{1}{N\left[1-f^c_N(k_1,\hdots,k_N)\right]^2}\sum_{i=1}^N
\nonumber\\
&&
\bigg\{\left[1-f^c_1(k_i)\right]^2f^c_{N-1}(\{k_j\}^{(N-1)}_{j\neq i})sG_1(k_i,s)
\nonumber \\
&&+\left[1-f^c_{N-1}(\{k_j\}^{(N-1)}_{j\neq i})\right]^2f^c_1(k_i)sG_{N-1}(\{k_j\}^{(N-1)}_{j\neq i},s)\bigg\}
\nonumber \\
&&=\frac{1}{N\left[1-f^c_N(k_1,\hdots,k_N)\right]^2}\sum_{i=1}^N
\left[1-f^c_1(k_i)\right]^2.
\nonumber \\
&&\times \left[1-f^c_{N-1}(\{k_j\}^{(N-1)}_{j\neq i})\right]^2
s\mathcal{L}\big[g_1(k_i,t)g_{N-1}(\{k_j\}^{(N-1)}_{j\neq i},t)\big]
\nonumber \\
\end{eqnarray}
In general, $G_n$ is a function of both $\epsilon$ and $s$ but for the critical point we set $\epsilon=0$. Hence the terms in the equations above are all integer powers of $sG_n(k_1,\hdots,k_n,s)$. These equations can be solved with the expansion 
\begin{gather}
G_n(k_1,\hdots,k_n,s)=\alpha_n(k_1,\hdots,k_n,s)
\nonumber\\
+\beta_n(k_1,\hdots,k_n,s)+\eta_n(k_1,\hdots,k_n,s)
\end{gather}
in which we make the ansatz $\lim_{s\rightarrow 0}\alpha_n(k_1,\hdots,k_n,s)/\beta_n(k_1,\hdots,k_n,s)=0$ 
and $\lim_{s\rightarrow 0}\beta_n(k_1,\hdots,k_n,s)/\eta_n(k_1,\hdots,k_n,s)=0$, i.e.\ we assume the terms to be of successively higher order, similar to the standard MCT approach.\cite{gotze1985properties}
In leading order we thus obtain $\bm{C}\bm{\alpha}=\bm{\alpha}$, which yields a solution $\alpha_n(k_1,\hdots,k_n,s)=\alpha(s)e_n(k_1,\hdots,k_n)$ with an undetermined factor $\alpha(s)$. Importantly, this $\alpha(s)$ is the same for all levels $n$, implying that all correlators $\phi_n(k_1,\hdots,k_n,t)$ relax in the same pattern as a function of time $t$. Physically, this can be understood from the fact that all correlators are explicit coupled within the GMCT hierarchy. 
Using Eq.~(\ref{eq:ehat}) to the next-to-leading order equations of Eqs.\ (\ref{eq:G_first}) and (\ref{eq:G_last}), we can determine $\alpha(s)$ via
\begin{equation}
\label{eq:alpha_critical}
s\alpha^2(s)=\lambda\mathcal{L}[\alpha(t)^2].
\end{equation}
This equation is exactly the same as the one in standard MCT,\cite{franosch1997asymptotic} except that the definition of $\lambda$ [Eq.~(\ref{eq:lambda})] now depends on the MF closure. 
Equation~(\ref{eq:alpha_critical}) can be solved by $\alpha(s)=\tilde{C}s^{x-1}\Gamma(1-x)$ or $\alpha(t)=\tilde{C}/t^x$, provided the exponent $x$ satisfies the equation $\Gamma(1-x)^2/\Gamma(1-2x)=\lambda$. Here $\tilde{C}$ is a constant related to the time scale which cannot be predicted in our analysis because Eq.~(\ref{eq:GMCT_laplace_3}) is time-scale invariant. There are two solutions for the exponent $x$, denoted as $a$ and $-b$. Thus the critical dynamics is 
\begin{equation}
\phi^c_n(k_1,\hdots,k_n,t)=f_n^c(k_1,\hdots,k_n)+h_n(k_1,\hdots,k_n)\left(\frac{t_0}{t}\right)^a
\label{eq:phia}
\end{equation} and 
\begin{equation}
\phi^c_n(k_1,\hdots,k_n,t)=f_n^c(k_1,\hdots,k_n)-h_n(k_1,\hdots,k_n)\left(\frac{t}{\tau}\right)^b,
\label{eq:phib}
\end{equation} where $t_0$ and $\tau$ are constants determined by different $\tilde{C}$'s. 
The time $t$ in Eq.~(\ref{eq:phia}) is in the range $\tau_\beta>t\gg t_0$ while in Eq.~(\ref{eq:phib}) $\tau_\beta<t\ll \tau$. 
Note that here the $n$-dependence is fully absorbed in $f^c_n$ and $h_n$, implying that the exponents $a$ and $b$ are the same for all levels.
This completes our analytical proof for the relation between $a$, $b$ and $\lambda$ and the power-law decay tested in the accompanying paper.\cite{luo2019generalized} 

\begin{figure}
	\epsfig{file=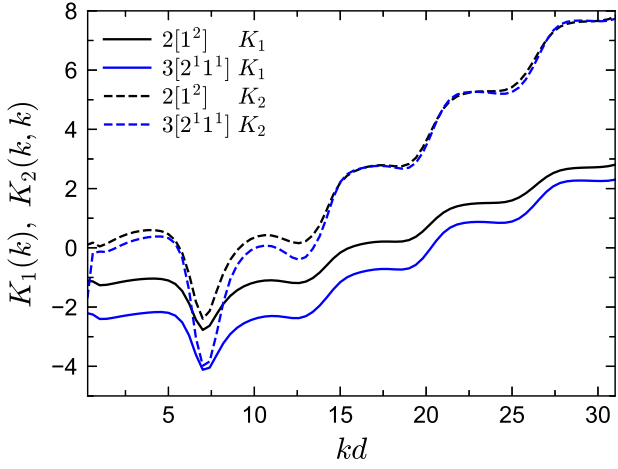,width=0.46\textwidth}
	\caption{\label{fig:Kn} 
		The amplitudes ${K}_1(k)$ and ${K}_2(k,k)$ as a function of wavenumber $k$ under different GMCT MF closure levels. 
                Solid lines are ${K}_1(k)$ and dashed lines are ${K}_2(k,k)$ under MF-$N[(N-1)^11^1]$ closure levels with $N=2,3$.              
	}
\end{figure}

In order to provide a valid time range for these scaling laws, we go to the next order $\beta_n(k_1,\hdots,k_n,s).$ Because the leading order of the right-hand side of  Eqs.\ (\ref{eq:G_first}) and (\ref{eq:G_last}) is $(sG_n)^2$, the order of $s\beta_n$ is $\left[s\alpha(s)\right]^2\sim s^{2x}$. Thus we can write $\beta_n(k_1,\hdots,k_n,s)=v_n(k_1,\hdots,k_n)\beta(s)$ with $\beta(s)=\tilde{C}^2s^{2x-1}\Gamma(1-2x).$
Again we separate $\bm{v}=\bm{v_0}+\kappa(x)\bm{e}$ with the special solution $\bm{v_0}=(\bm{1}-\bm{C})^{-1}\bm{R}$ but now the $\bm{R}$ contains only two terms, 
\begin{eqnarray*}
R_n(k_1,\hdots,k_n)=-\lambda\left[1-f_n^c(k_1,\hdots,k_n)\right]e_n^2(k_1,\hdots,k_n), 
\\
\hspace{1mm}\text{for} \ \ \ n<N
\end{eqnarray*}
and
\begin{gather}
R_{N}(k_1,\hdots,k_N)=\frac{1}{\left[1-f^c_N(k_1,\hdots,k_N)\right]^2}\frac{1}{N}\sum_{i=1}^N
\nonumber \\ 
\left[1-f^c_1(k_i)\right]^2\left[1-f^c_{N-1}(\{k_j\}^{(N-1)}_{j\neq i})\right]^2e_1(k_i)e_{N-1}(\{k_j\}^{(N-1)}_{j\neq i}).
\end{gather}
The $\kappa(x)$ (where $x$ can be $a$ or $-b$) can be calculated similarly as in the long-time limit section, namely by solving Eqs.~(\ref{eq:G_first}) and (\ref{eq:G_last}) up to the higher order $s\eta_n(s)\sim \left[s\alpha(s)\right]^3\sim s^{3x}$ with Eq.~(\ref{eq:ehat}):
\begin{gather}
\kappa(x)=\frac{\xi\Gamma(1-3x)-\zeta\Gamma^3(1-x)}{\Gamma(1-x)\Gamma(1-2x)-\lambda\Gamma(1-3x)},
\end{gather}
where 
\begin{gather}
\xi=\frac{1}{2}\sideset{}{'}\sum_{N}\hat{e}_N(k_1,\hdots,k_N)\frac{1}{\left[1-f^c_N(k_1,\hdots,k_N)\right]^2}\frac{1}{N}
\nonumber \\
\times \sum_{i=1}^N\left[1-f^c_1(k_i)\right]^2\left[1-f^c_{N-1}(\{k_j\}^{(N-1)}_{j\neq i})\right]^2
\nonumber \\
\times \left[e_1(k_i)v_{0N-1}(\{k_j\}^{(N-1)}_{j\neq i})+v_{01}(k_i)e_{N-1}(\{k_j\}^{(N-1)}_{j\neq i})\right]
\end{gather}
and 
\begin{gather}
\zeta=\sideset{}{'}\sum_{n<N}\hat{e}_n(k_1,\hdots,k_n)\big\{
\nonumber \\
\frac{1}{2}\left[1-f^c_n(k_1,\hdots,k_n)\right]^2e^{3}_n(k_1,\hdots,k_n)
\nonumber \\
+\frac{1}{\lambda}\left[1-f^c_n(k_1,\hdots,k_n)\right]e_n(k_1,\hdots,k_n)v_{0n}(k_1,\hdots,k_n)\big\}.
\end{gather}
Finally we obtain the leading preasymptotic correction of the critical decay 
\begin{gather}
\phi^c_n(k_1,\hdots,k_n,t)=f_n^c(k_1,\hdots,k_n)+
\nonumber \\
h_n(k_1,\hdots,k_n)\left(\frac{t_0}{t}\right)^a\left\{1+\left[K_n(k_1,\hdots,k_n)+\kappa(a)\right]\left(\frac{t_0}{t}\right)^a\right\}
\label{eq:criticala}
\end{gather} and 
\begin{gather}
\phi^c_n(k_1,\hdots,k_n,t)=f_n^c(k_1,\hdots,k_n)-
\nonumber\\
h_n(k_1,\hdots,k_n)\left(\frac{t}{\tau}\right)^b\left\{1-\left[K_n(k_1,\hdots,k_n)+\kappa(-b)\right]\left(\frac{t}{\tau}\right)^b\right\},
\label{eq:criticalb}
\end{gather}
where 
\begin{equation}
K_n(k_1,\hdots,k_n)=\frac{v_{0n}(k_1,\hdots,k_n)}{e_n(k_1,\hdots,k_n)}.
\end{equation}
The leading-order results of Eq.~(\ref{eq:phia}) and Eq.~(\ref{eq:phib}) are thus applicable only when $t/t_0\gg \left[{K}_n(k_1,\hdots,k_n)+\kappa(a)\right]^{1/a}$ and $t/\tau\ll {\left[{K}_n(k_1,\hdots,k_n)+\kappa(-b)\right]^{-1/b}}$, respectively.
Notably, Eq.~(\ref{eq:criticalb}) is the so-called von Schweidler law.

\begin{figure}
	\epsfig{file=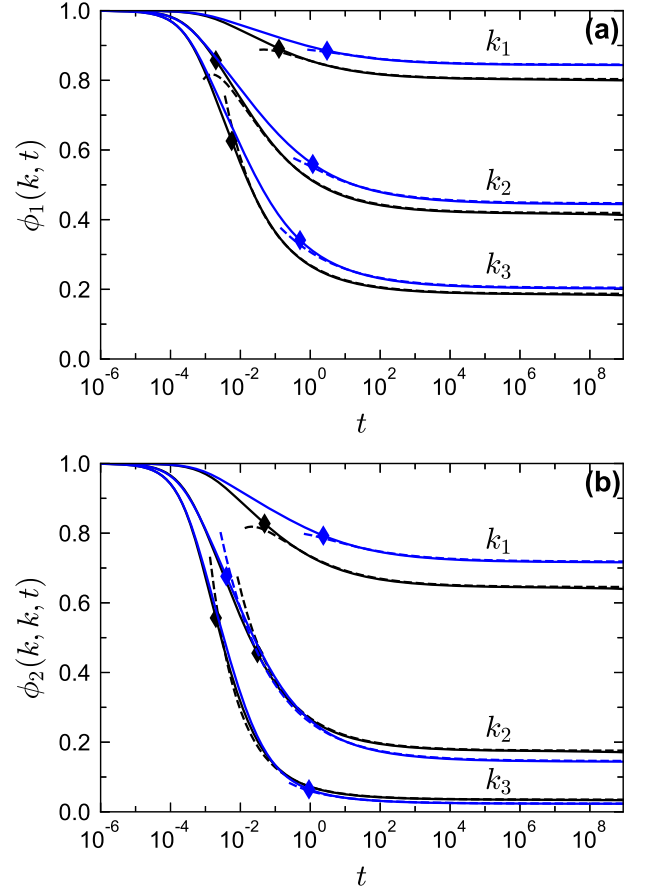,width=0.46\textwidth}
	\caption{\label{fig:criticalphi} 
		The time-dependent density correlation functions $\phi_1(k,t)$ and $\phi_2(k,k,t)$ at critical point $\varphi=\varphi^c$ under MF-$N[(N-1)^11^1]$ closures with $N=2$ (black lines) and $N=3$ (blue lines).
                Solid lines are the numerical solutions and dashed lines are the leading and next-to-leading asymptotic results using Eq.\ (\ref{eq:criticala}).           The three wavenumbers are the same as those in Fig.~\ref{fig:f}. Diamonds marks the point with relative error $10\%$ for $\phi_1(k,t)-f_1^c(k)$ and $\phi_2(k,k,t)-f_2^c(k,k)$ in panel \textbf{(a)} and \textbf{(b)}, respectively. The fitting time parameter is  $t_0\approx 3.3 \times10^{-4}$ for both closure levels $N=2$ and $N=3$.
	}
\end{figure}

We have already numerically tested the leading order solutions Eqs.~(\ref{eq:phia}) and (\ref{eq:phib}) for $\phi_1(k,t)$ in the accompanying paper.\cite{luo2019generalized} Here, we further test the leading preasymptotic correction of the first step decay [Eq.~(\ref{eq:criticala})] for PY hard spheres;
the test of the second step decay (i.e.\ the von Schweidler dynamics) will be treated in the next section. 
Figure \ref{fig:Kn} show the amplitudes $K_1(k)$ and $K_2(k,k)$ under MF-$N[(N-1)^11^1]$ closure levels with $N=2,3$. The shapes of $K_1(k)$ and $K_2(k,k)$ are similar to the $\bar{K}_1(k)$ and $\bar{K}_2(k,k)$ [see Eq.\ (\ref{eq:Kbar}) and Fig.\ \ref{fig:Knbar}]. Moreover, as shown in Fig.~\ref{fig:criticalphi}, Eq.~(\ref{eq:criticala}) can accurately describe the critical decay over more than eight decades in time for all wavenumbers and all closure levels considered here. We thus conclude that, within the context of GMCT, our asymptotic results with leading corrections provide a good analytical description of the critical decay at the glass transition point for arbitrary-order density correlators. 

\begingroup
\setlength{\tabcolsep}{8.3pt} 
\renewcommand{\arraystretch}{1.1} %
\begin{table*}
	\caption{Predicted critical packing fractions $\varphi^c$ and parameters for Percus-Yevick hard spheres obtained 
		under different GMCT MF-$N[(N-1)^11^1]$ levels. For level $N=4$, the coefficient matrix is too large to obtain the second order parameters $\kappa$, $\xi$ and $\zeta$ within reasonable computing time.
	}
	\begin{tabular}{lccccccccccc} 
		\hline
		\hline
		\\[-1em]
		MF level & $\varphi^c$&$\gamma$&$a$&$b$&$\lambda$&$\sigma$&$\kappa$&$\xi$&$\zeta$&$\kappa(a)$&$\kappa(-b)$\\
		\\[-1em]
		\hline
		\\[-1em]
		$2[1^2]$ &0.515914 &2.4544&0.3124&0.5856&0.7334&0.386&2.08&-0.227&-0.461&0.298&0.181\\
		$3[2^11^1]$ & 0.531888 &2.7110&0.2895&0.5083&0.7804&0.346&3.46&-0.3244&-0.9756&0.0915&-1.4769 \\
		$4[3^11^1]$ & 0.544172 &2.9531&0.2707&0.4521&0.8143&0.289\\
		\\[-1em]
		\hline
		\hline	
	\end{tabular}
	\label{tab:coeff}
\end{table*}
\endgroup


\section{$\beta$ scaling-law regime}
Let us now consider the scaling laws in the $\beta$-regime where $|\phi_n(t)-f^c|\ll 1$ for $\epsilon\neq0$. In this case Eq.~(\ref{eq:G_first}) and (\ref{eq:G_last}) are still applicable and both $|\epsilon|$ and $g_n(k_1,\hdots,k_n,t)$ $\big[$or $sG_n(k_1,\hdots,k_n,s)\big]$  can be treated as small quantities. We expand $G_n(k_1,\hdots,k_n,s)$ in powers of $\sqrt{|\epsilon|}$
\begin{gather}
G_n(k_1,\hdots,k_n,s)=\alpha_n(k_1,\hdots,k_n,s)\sqrt{|\epsilon|}
\nonumber \\
+\beta_n(k_1,\hdots,k_n,s)|\epsilon|+\eta_n(k_1,\hdots,k_n,s)|\epsilon|^{3/2}.
\end{gather}


To order $\sqrt{|\epsilon|}$, we obtain $\alpha_n(k_1,\hdots,k_n,s)=e_n(k_1,\hdots,k_n)\alpha(s)$. To order $|\epsilon|$, however, we obtain a different equation to determine $\alpha(s)$ compared to the critical case, 
\begin{equation}
\label{eq:alpha_sigma}
\sigma-s^2\alpha^2(s)+\lambda s \mathcal{L}[\alpha(t)^2]=0.
\end{equation}
Again, this equation is the same as the corresponding one in standard MCT.\cite{franosch1997asymptotic}
When $\epsilon$ tends to zero, the solution should approach the one at the critical point [Eq.~(\ref{eq:phia})]. Thus $\lim_{|\epsilon|\to0}\sqrt{|\epsilon|}\alpha(t)(\frac{t}{t_0})^a=1$, from which 
we obtain the time scale characterizing the $\beta$-relaxation regime
\begin{equation}
\tau_\beta=t_0\left(\sigma|\epsilon|\right)^{-\frac{1}{2a}}\sim |\epsilon|^{-\frac{1}{2a}}.
\label{eq:tau_beta}
\end{equation}
We also obtain that  
\begin{equation}
\alpha(\hat{t})=\sqrt{\sigma}g_{\pm}(\hat{t}), \ \ \hat{t}=t/\tau_\beta
\end{equation} 
where the $g_{\pm}$ define the solutions of Eq.~(\ref{eq:alpha_sigma}) for $\sigma=\pm1$.
Since Eq.~(\ref{eq:alpha_sigma}) is exactly same as in standard MCT,\cite{franosch1997asymptotic} the obtained $g_\pm(t)$ are also same as those in standard MCT. Specfically, for small rescaled times $\hat{t}=t/\tau_\beta$, 
\begin{eqnarray}
g_\pm(\hat{t})=1/\hat{t}^a\pm A_1\hat{t}^a+O(\hat{t}^{3a}),
\nonumber \\
A_1=\frac{1}{2[\Gamma(1+a)\Gamma(1-a)-\lambda]}.
\label{eq:gpm}
\end{eqnarray}
For large  $\hat{t}=t/\tau_\beta\gg 1$,
\begin{eqnarray}
g_-(\hat{t})=-B\hat{t}^b+ B_1/(B\hat{t}^b)+O(1/\hat{t}^{3b}),
\nonumber \\
B_1=\frac{1}{2[\Gamma(1-b)\Gamma(1+b)-\lambda]},
\label{eq:gminus}
\end{eqnarray}
where $B$ depends only on $\lambda$ but has to be determined from matching the asymptotic solution.\cite{Gotze1990}
Therefore, to leading order $\sqrt{|\epsilon|}$, the scaling law of the $\beta$-relaxation regime reads
\begin{gather}
\phi_n(k_1,\hdots,k_n,t)=f_n^c(k_1,\hdots,k_n)
\nonumber \\
+h_n(k_1,\hdots,k_n)\sqrt{\sigma|\epsilon|} g_\pm(\hat{t})
\label{eq:beta1}
\end{gather}


To calculate the leading corrections of the scaling laws to order $|\epsilon|^{3/2}$,  we determine $\beta_n(k_1,\hdots,k_n,t)$ via
\begin{gather}
\beta_n(k_1,\hdots,k_n,t)=e_n(k_1,\hdots,k_n)\left[h(t)+\chi\right]
\nonumber \\
+v_{0n}(k_1,\hdots,k_n)\left[\alpha^2(t)-\frac{\sigma}{1-\lambda} \right]+B_{0n}(k_1,\hdots,k_n).
\end{gather}
Here, $h(t)$ is the correction-to-scaling master function\cite{gotze1989beta} determined from the order $\epsilon^{3/2}$ in $G_n$ via Eq.~(\ref{eq:ehat}),
\begin{equation}
s\alpha(s)h(s)-\lambda\mathcal{L}\left[\alpha(t)h(t)\right]=\xi\mathcal{L}\left[\alpha^3(t)\right]-\lambda\zeta s\alpha(s) \mathcal{L}\left[\alpha^2(t)\right].
\label{eq:ht}
\end{equation}
The constant $\chi$ can also be calculated from the order $\epsilon^{3/2}$ simultaneously,
 \begin{equation}
  \chi=\frac{\sigma(\lambda\zeta-\xi)}{(1-\lambda)^2}+\sqrt{\frac{\sigma}{1-\lambda}}\kappa.
 \end{equation} 
Equation~(\ref{eq:ht}) is the same as the one in standard MCT and the properties of $h(t)$ were already well studied.\cite{franosch1997asymptotic} 
Overall, the $\beta$ scaling expression for $\phi_n(k_1,\hdots,k_n,t)$ including next-to-leading order corrections is 
\begin{gather}
\phi_n(k_1,\hdots,k_n,t)=f_n^c(k_1,\hdots,k_n)
\nonumber\\
+h_n(k_1,\hdots,k_n)\bigg\{\sqrt{|\epsilon|}\alpha(t)+|\epsilon|h(t)+\epsilon \chi
\nonumber\\
+\bigg[K_n(k_1,\hdots,k_n)\left(|\epsilon|\alpha^2(t)-\epsilon\frac{\sigma}{1-\lambda} \right)
\nonumber\\
+\epsilon\sqrt{\frac{\sigma}{1-\lambda}}\bar{K}_n(k_1,\hdots,k_n)\bigg]\bigg\}.
\end{gather}
Notice that the correction $|\epsilon|h(t)+\epsilon\chi$ does not lead to a violation of the factorization theorem. 
Using the properties of $g_-(t)$ and $h_-(t)$ \cite{franosch1997asymptotic} we obtain a simpler expression for large $\hat{t}$,
\begin{gather}
\phi_n(k_1,\hdots,k_n,t)=f_n^c(k_1,\hdots,k_n)
\nonumber\\
-h_n(k_1,\hdots,k_n)\left(\frac{t}{\tau}\right)^b\left\{1-\left[K_n(k_1,\hdots,k_n)+\kappa(-b)\right]\left(\frac{t}{\tau}\right)^b\right\},
\label{eq:beta2}
\end{gather}
which recovers the von Schweidler law of Eq.~(\ref{eq:criticalb}) in the critical case ($\epsilon\rightarrow0^-$), as expected.
\begin{figure}
	\epsfig{file=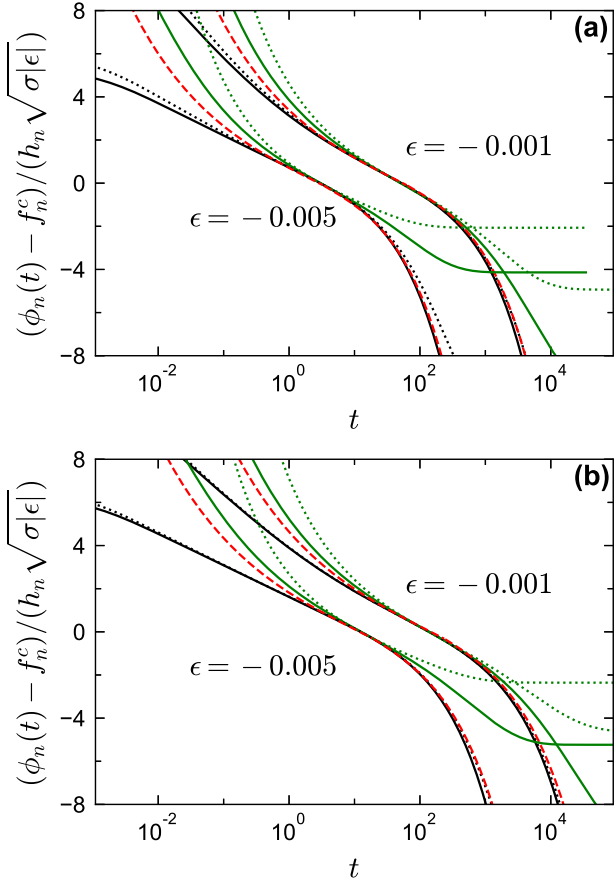,width=0.46\textwidth}
	\caption{\label{fig:g} 
	Test of $g_{-}(\hat{t})$ for the $\beta$-relaxation scaling laws. \textbf{(a)} The evolution of $(\phi_1(k,t)-f^c_1(k))/(h_1(k)\sqrt{\sigma|\epsilon|})$ and $(\phi_2(k,k,t)-f^c_2(k,k))/(h_2(k,k)\sqrt{\sigma|\epsilon|})$ as a function of time $t$ for two different reduced packing fractions $\epsilon=-0.001$ and $\epsilon=-0.005$ under closure MF-$2[(2-1)^11^1]$, i.e.\ the standard MCT case. For each $\epsilon$, two wavenumbers $k_1d=7.4$ (black lines) and $k_3d=17.4$ (green lines) are plotted. The solid lines and dotted lines corresponds to $n=1$ and $n=2$, respectively. The red dashed lines are the $g_-(\hat{t})$ from Eq.~(\ref{eq:gpm}) and (\ref{eq:gminus}). For clarity, we do not scale $\phi_n(t)$ over $\tau_\beta$ but rather plot $g_-(t/\tau_\beta)$ as a function of $t$. \textbf{(b)} Same as (a) but under closure MF-$3[(3-1)^11^1]$. 
	}
\end{figure}

\begin{figure}
	\epsfig{file=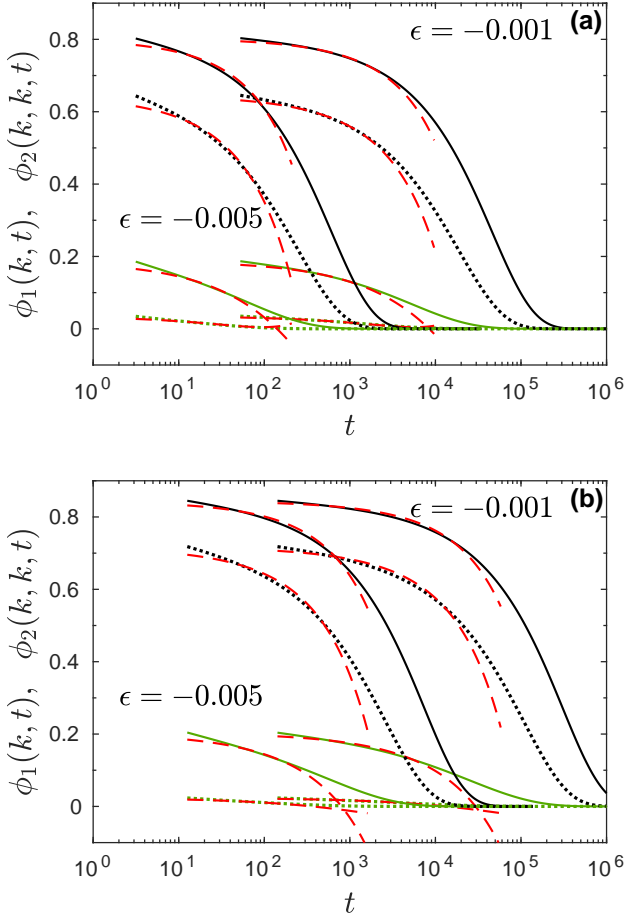,width=0.46\textwidth}
	\caption{\label{fig:latebeta} 
	Test of $\beta$-relaxation scaling laws at late $t/\tau_\beta$. \textbf{(a)} The evolution of $\phi_1(k,t)$ (solid lines) and $\phi_2(k,k,t)$ (dotted lines) as a function of time $t$ for two different reduced packing fractions $\epsilon=-0.001$ and $\epsilon=-0.005$ under closure MF-$2[(2-1)^11^1]$. For each $\epsilon$, two wavenumbers $k_1d=7.4$ (black lines) and $k_3d=17.4$ (green lines) are plotted. The red dashed lines are the results from Eq.~(\ref{eq:beta2}). \textbf{(b)} Same as (a) but under closure MF-$3[(3-1)^11^1]$. 
	}
\end{figure}

Next we test Eqs.\ (\ref{eq:beta1}) and (\ref{eq:beta2}) numerically, as has also been widely done in MCT, especially when comparing with simulations or experiments.  For Eq.~(\ref{eq:beta1}), the scaling behavior of $\phi_1(k,t)$ over wavenumbers and $\epsilon$ was already tested in the accompanying paper.\cite{luo2019generalized} We further test Eq.~(\ref{eq:beta1}) for PY hard spheres with the expressions $g_{-}(\hat{t})$ in Eqs.~(\ref{eq:gpm}) and (\ref{eq:gminus}) for both $\phi_1(k,t)$  and $\phi_2(k,k,t)$. We show this scaling law and the $g_-(\hat{t})$ under closure MF-$N[(N-1)^11^1]$ when $N=2$ [Fig.~\ref{fig:g}(a)] and $N=3$ [Fig.~\ref{fig:g}(b)]. It is clear that both $\phi_1(k,t)$  and $\phi_2(k,k,t)$ satisfy the scaling law, as the solid lines and dotted lines collapse. This confirms our results that all correlators $\phi_n(k_1,\hdots,k_n,t)$ relax in a same pattern as a function of $t$, in particular demonstrating that all correlators reach their respective plateaus, $f^c_n(k_1,\hdots,k_n)$,  simultaneously.  Moreover, all lines collapse to the master curve $g_-(\hat{t})$ (red dashed lines in Fig.~\ref{fig:g}). Although the valid time ranges are different for different wavenumbers (compare the black lines and green lines in Fig.~\ref{fig:g}) and the absolute time range decreases when $|\epsilon|$ increases (compare the lines for $\epsilon=-0.001$ and $\epsilon=-0.005$ in Fig.~\ref{fig:g}), these phenomena are similar for both closure levels $N=2$ and $N=3$. This shows that with respect to the scaling behavior, there is not much difference when the closure level $N$ increases. Therefore, Eqs.~(\ref{eq:beta1}),  (\ref{eq:gpm}), and (\ref{eq:gminus}) can successfully describe the dynamics for higher-order density correlation functions in the $\beta$-relaxation regime near the corresponding plateaus, i.e.\ where the slowdown due to the cage effect is manifested most markedly.
 Finally, in Fig.~\ref{fig:latebeta} we plot the predictions from Eq.~(\ref{eq:beta2}) in the late $\beta$-relaxation regime, i.e.\ the dynamics of the correlators for times beyond $\tau_\beta$.
 Similar to the above case, we find that the von Schweidler law is also applicable for both MF closure level $N=2$ [Fig.~\ref{fig:latebeta}(a)] and $N=3$ [Fig.~\ref{fig:latebeta}(b)] and for all wavenumbers (black lines and green lines in Fig.~\ref{fig:latebeta}). 
 
 In sum, we have analytically obtained the leading and next-to-leading order expressions for the dynamics of multi-point density correlation functions in the $\beta$-relaxation regime, as given by Eqs.\  (\ref{eq:beta1}) and (\ref{eq:beta2}). These scaling laws, which we also confirm numerically, are among the most representative triumphs of MCT; here we find that they can also be generalized to the broader GMCT framework.

\section{$\alpha$ scaling-law regime}
Finally, we consider the relaxation dynamics in the $\alpha$ scaling-law regime. This regime emerges for $\varphi<\varphi^c$ as a second relaxation step towards equilibrium, and deals with the dynamics on the time scale $\tau$. We first establish the power-law divergence of $\tau$ with $\epsilon$ for GMCT under MF closures, consistent with the well-known standard MCT result. After a reformulation of the equations of motion, we will then establish the general existence of a time-density superposition principle for the $\alpha$ process, which has been numerically tested in the accompanying paper.\cite{luo2019generalized}


We first relate the results in the $\beta$-scaling-law regime to the $\alpha$-relaxation time $\tau$ by comparing Eq.~(\ref{eq:phib}) and the large $\hat{t}$ limit of Eqs.~(\ref{eq:gminus}) and (\ref{eq:beta1}) such that $\sqrt{\sigma|\epsilon|}B\hat{t}^b=(t/\tau)^b$. Combined with Eq.~(\ref{eq:tau_beta}), this leads to the power law of the $\alpha$-relaxation time 
\begin{equation}
\tau=\tau_{\beta}B^{-1/b}\left(\sigma|\epsilon|\right)^{-\frac{1}{2b}}=t_0B^{-1/b}\left(\sigma|\epsilon|\right)^{-\left(\frac{1}{2a}+\frac{1}{2b}\right)}\sim |\epsilon|^{-\gamma},
\end{equation}
where $\gamma=\left(\frac{1}{2a}+\frac{1}{2b}\right)$. Thus, even though the values of the exponents $a$, $b$, and $\gamma$ quantitatively change with GMCT closure level $N$, their non-trivial connection remains the same for all levels. 

Let us now introduce rescaled times $\tilde{t}=t/\tau$ and $\tilde{s}=s\tau$. We carry out an asymptotic expansion for small negative reduce packing fraction $\epsilon$,
\begin{gather}
\phi_n(k_1,\hdots,k_n,t)=\tilde{\phi}_n(k_1,\hdots,k_n,\tilde{t})
\nonumber \\
+\epsilon\left[1-f^c_n(k_1,\hdots,k_n)\right]^2\tilde{\psi}_n(k_1,\hdots,k_n,\tilde{t})+O(\epsilon^2),
\end{gather}
and the corresponding frequency form
\begin{gather}
s\Phi_n(k_1,\hdots,k_n,s)=\tilde{s}\tilde{\Phi}_n(k_1,\hdots,k_n,\tilde{s})
\nonumber \\
+\epsilon\left[1-f^c_n(k_1,\hdots,k_n)\right]^2\tilde{s}\tilde{\Psi}_n(k_1,\hdots,k_n,\tilde{s})+O(\epsilon^2),
\end{gather}
which satisfy Eq.~(\ref{eq:GMCT_laplace_3}) and Eq.~(\ref{eq:closure_laplace}) with rescaled $\tilde{t}$ and $\tilde{s}$. Specializing to $\epsilon=0$ we obtain the equation for the leading-order contribution $\tilde{\Phi}_n(k_1,\hdots,k_n,\tilde{s})$
\begin{gather}
\frac{\tilde{s}\tilde{\Phi}_n(k_1,\hdots,k_n,\tilde{s})}{1-\tilde{s}\tilde{\Phi}_n(k_1,\hdots,k_n,\tilde{s})}=
\nonumber \\
=\sum_{iqp}V^c_{n+1}(k_1,\hdots ,k_n,q,p,i)\tilde{s}\tilde{\Phi}_{n+1}(q,p,\{k_j\}^{(n-1)}_{j\neq i},\tilde{s}),
\end{gather}
and
\begin{gather} 
\tilde{s}\tilde{\Phi}_N(k_1,\hdots,k_N,\tilde{s})=
\nonumber \\
\frac{1}{N}\sum_{i=1}^N \tilde{s}\mathcal{L}\big[\tilde{\phi}_1(k_i,\tilde{t})\times \tilde{\phi}_{N-1}(\{k_j\}^{(N-1)}_{j\neq i},\tilde{t})\big].
\end{gather}
Indeed, these equations describe the dynamics at the critical point, hence for small $\tilde{t}$ the solutions are identical to Eq.~(\ref{eq:criticalb}). We thus find that to leading order we can write  ${\phi}_n(k_1,\hdots,k_n,{t})\approx\tilde{\phi}_n(k_1,\hdots,k_n,\tilde{t})$, which constitutes the so-called time-density or time-temperature superposition principle. More explicitly, after rescaling the time $t$ with a density- or temperature-dependent $\tau$, the dynamics should conform to the same master curve. 
\begin{figure}
	\epsfig{file=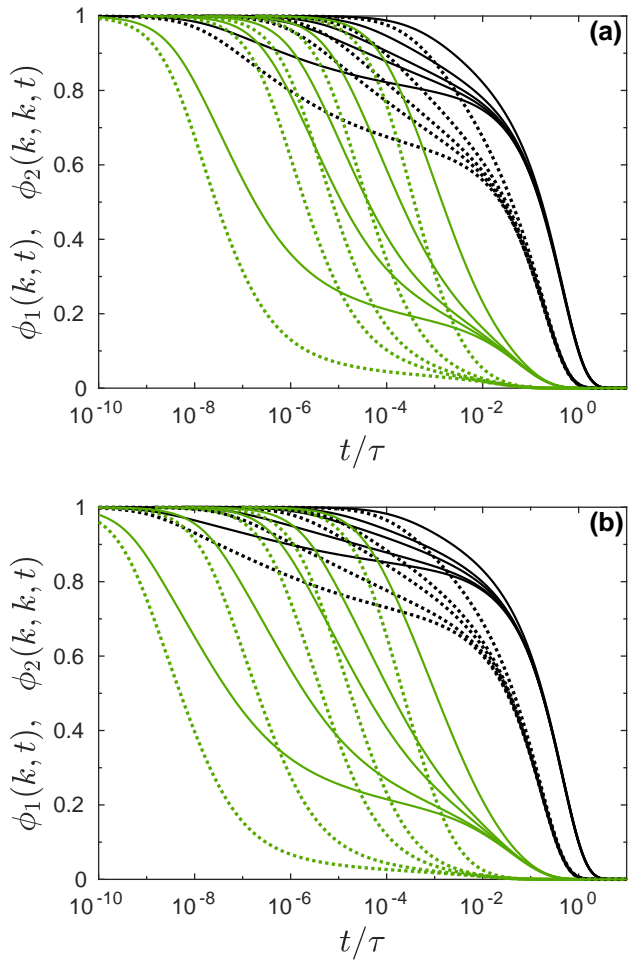,width=0.46\textwidth}
	\caption{\label{fig:alpha} 
	Test of $\alpha$-relaxation scaling laws. \textbf{(a)} The evolution of $\phi_1(k,t)$ (solid lines) and $\phi_2(k,k,t)$ (dotted lines) as a function of time $t/\tau$ under closure MF-$2[(2-1)^11^1]$. For each $\epsilon$, two wavenumbers $k_1d=7.4$ (black lines) and $k_3d=17.4$ (green lines) are plotted. From left to right the $\epsilon$ are $-0.001$, $-0.005$, $-0.01$, $-0.02$, $-0.05$. \textbf{(b)} Same as (a) but under closure MF-$3[(3-1)^11^1]$. 
	}
\end{figure}
If we use the same degree of accuracy for the description of the $\alpha$ process as the leading contribution to the  $\beta$ scaling laws, i.e.\ up to order $|\epsilon|^{1/2}$, we only need the above solution since there are no corrections to the $\alpha$ scaling-law results of order $|\epsilon|^{1/2}$. This observation implies that the superposition principle also holds for larger $\epsilon$, where the $\beta$ scaling-law results are no longer applicable.

Considering the order $\epsilon$ of Eq.~(\ref{eq:GMCT_laplace_3}) and Eq.~(\ref{eq:closure_laplace}), the leading correction $\tilde{\psi}_n(k_1,\hdots,k_n,\tilde{t})$ satisfies
\begin{gather}
\frac{\left[1-f_n^c\left(k_1,\hdots,k_n\right)\right]^2\tilde{s}\tilde{\Psi}_n(k_1,\hdots,k_n,\tilde{s})}{\left[1-\tilde{s}\tilde{\Phi}_n(k_1,\hdots,k_n,\tilde{s})\right]^2}=
\nonumber \\
=\sum_{iqp}V^{'c}_{n+1}(k_1,\hdots ,k_n,q,p,i)\tilde{s}\tilde{\Phi}_{n+1}(q,p,\{k_j\}^{(n-1)}_{j\neq i},\tilde{s})
\nonumber\\
+\sum_{iqp}C^{c}_{n+1}(k_1,\hdots ,k_n,q,p,i)\tilde{s}\tilde{\Psi}_{n+1}(q,p,\{k_j\}^{(n-1)}_{j\neq i},\tilde{s})
\nonumber \\
\hspace{50mm}\text{for} \ \ \ n<N
\end{gather}
and 
\begin{gather} 
\tilde{s}\tilde{\Psi}_N(k_1,\hdots,k_N,\tilde{s})=
\frac{1}{N\left[1-f_N^c\left(k_1,\hdots,k_N\right)\right]^2}\times\sum_{i=1}^N \tilde{s}\mathcal{L}\bigg\{
\nonumber \\
\tilde{\phi}_1(k_i,\tilde{t})\times\left[1-f_{N-1}^c\left(\{k_j\}^{(N-1)}_{j\neq i}\right)\right]^2 \tilde{\psi}_{N-1}(\{k_j\}^{(N-1)}_{j\neq i},\tilde{t})
\nonumber \\
+\tilde{\psi}_1(k_i,\tilde{t})\times\left[1-f_1^c\left(k_i\right)\right]^2 \tilde{\phi}_{N-1}(\{k_j\}^{(N-1)}_{j\neq i},\tilde{t})\bigg\}.
\end{gather}
These equations can be solved for small $\tilde{t}$ by an expansion in powers of $\tilde{t}^b$; this yields the following $\alpha$-scaling expression for $\phi_n(k_1,\hdots,k_n,t)$ including next-to-leading order corrections:
\begin{gather}
\phi_n(k_1,\hdots,k_n,t)=
\nonumber \\
\tilde{\phi}_n(k_1,\hdots,k_n,\tilde{t})+h_n(k_1,\hdots,k_n)B_1\sigma\epsilon\tilde{t}^{-b},
\label{eq:alpha_last}
\end{gather}
with the $B_1$ defined in Eq.~(\ref{eq:gminus}) and the $h_n(k_1,\hdots,k_n)$ in Eq.~(\ref{eq:hn}).

To numerically test the validity of the time-density superposition principle [Eq.~(\ref{eq:criticalb})] for PY hard spheres, we show the relaxation of density correlation functions as a function of the rescaled time $t/\tau$ in Fig.~\ref{fig:alpha}. We plot the relaxations of $\phi_1(k,t)$ and $\phi_2(k,k,t)$ at five different values of $\epsilon$, from $-0.001$ to $-0.05$, at wavenumbers $k_1d=7.4$ and $k_3d=17.4$, under MF closure levels $N=2$ and $N=3$. For a given $\epsilon$, the $\tau$ we used for $\phi_1(k,t)$ and $\phi_2(k,k,t)$ are the same at both wavenumbers. We checked that different definitions of $\tau$ such as $\phi_1(k_1,\tau)=0.1$, $\phi_1(k_1,\tau)=e^{-1}$ or $\phi_1(k_1,\tau)=0.1\times f^c_1(k_1)$, all give a robust power-law behavior $\tau\sim\epsilon^{-\gamma}$. It can be seen that all curves at different $\epsilon$ collapse fairly well, which means this superposition principle is indeed also applicable for higher-order multi-point density correlation functions and, importantly, all of them have a similar $\alpha$-relaxation time $\tau\sim\epsilon^{-\gamma}$. This further corroborates our conclusion that all multi-point correlators, at different levels and different wavenumbers, decay in the same main pattern. However, we can see that at $k_3d$, the deviation seems larger compared with $k_1d$. This is due to the correction term up to order $\epsilon$, Eq.~(\ref{eq:alpha_last}), leading to a relative error proportional to the ratio $h_1(k)/f_1^c(k)$ or $h_2(k,k)/f_2^c(k,k)$ for $\phi_1(k,t)$ or $\phi_2(k,k,t)$, respectively, similar to MCT.\cite{franosch1997asymptotic} From Fig.~\ref{fig:fc} and Fig.~\ref{fig:h} we know that at $k_1d$ the ratio is smaller than the other wavenumbers, which explains why the scaling law usually works better around the peak of $S(k)$. Again, for GMCT closure levels $N=2$ and $N=3$, the scaling behaviors are almost the same. 
\begingroup
\setlength{\tabcolsep}{7.pt} 
\renewcommand{\arraystretch}{1.1} %
\begin{table}
	\caption{Predicted critical packing fractions $\varphi^c$ and parameters $\gamma$, $a$, $b$, $\lambda$ for Percus-Yevick hard spheres obtained 
		under MF-$N[1^N]$ closures.
	}
	\begin{tabular}{lccccc} 
		\hline
		\hline
		\\[-1em]
		MF level & $\varphi^c$&$\gamma$&$a$&$b$&$\lambda$\\
		\\[-1em]
		\hline
		\\[-1em]
		$3[1^3]$ & 0.526624 &2.5792&0.3008&0.5452&0.7579\\
		$4[1^4]$ & 0.535382 &2.7094&0.2896&0.5087&0.7801\\
		\\[-1em]
		\hline
		\hline	
	\end{tabular}
	\label{tab:coeff2}
\end{table}
\endgroup

\section{Conclusions}
In this work, we have presented the asymptotic solutions and some leading preasymptotic corrections for structural relaxation in the vicinity of the glass transition within first-principles-based generalized mode-coupling theory. 
Our results, which generalize the well-established scaling laws of MCT to multi-point density correlations, are in good agreement with numerical data, and may be extended to arbitrary-order density correlation functions under arbitrary GMCT mean-field closures. In our derivations the only assumption we have made use of relies on the property that the static structure factor changes almost linearly when the density (or temperature) shifts by a small value away from the critical point. This property is ubiquitous, recalling that the main challenge in the field is indeed to predict the dramatic dynamical slowdown from only minor changes in the structure. We therefore expect our solutions to be generally applicable for  glass-forming materials close to the critical point, but the density or temperature applicability ranges of these solutions may be material-dependent. 

The analytical solutions for the glass form factors near the critical point are described by Eq.~(\ref{eq:long_time_asy}) and they have also been verified numerically for the PY hard-sphere system. Although at first glance our expression for arbitrary-order GMCT looks similar to the result of standard MCT, there is an important physical difference emerging from the present hierarchical GMCT analysis. 
In view of the non-trivial trend of the amplitude $h_2(k,k)$ and $\bar{K}_2(k,k)$ with closure level $N$, as well as the inequality $[f_1^c(k)]^2\neq f_2^c(k,k)$ that we find for $N>2$, we hypothesize that GMCT may also account at least in part for dynamically heterogeneous dynamics. This is encouraging, as conventional MCT is known to neglect the many-body spatiotemporal density fluctuations underlying dynamical heterogeneity by virtue of the MCT approximation  $[f_1^c(k)]^2\equiv f_2^c(k,k)$. Future work should establish to which extent higher-order GMCT can quantitatively capture dynamical heterogeneities and activated dynamics on a strictly first-principles basis. 

We have also derived the asymptotic and preasymptotic solutions for the time-dependent multi-point density correlation functions. At the glass transition point, the leading order solutions for the critical decay and the von Schweidler dynamics are given by Eqs.~(\ref{eq:phia}) and (\ref{eq:phib}), respectively, and the solutions with leading corrections are given by Eqs.~(\ref{eq:criticala}) and (\ref{eq:criticalb}). These regimes correspond to the early and late $\beta$ process, respectively, which are separated by the $\beta$-relaxation time scale $\tau_\beta$. The non-trivial relation between the power-law exponents $a$, $b$, and $\lambda$, already established for standard MCT, also emerges from our GMCT derivation under mean-field closures. For the supercooled-liquid phase near the critical point, we have also derived the general $\beta$-regime scaling laws. Here we find that the leading-order master functions $g_{\pm}(\hat{t})$ satisfy the same wavenumber-independent evolution equation as expected from standard MCT. Our $\beta$-relaxation time scale $\tau_\beta$ grows as a power law $\tau_\beta\sim\epsilon^{-1/2a}$ and the $\alpha$-relaxation time scale $\tau$ grows as a power law $\tau\sim\epsilon^{-\gamma}$, again fully consistent with the well-known MCT results. Moreover, we analytically confirm the existence of the time-density (or time-temperature) superposition principle in the $\alpha$-relaxation regime. Overall, we conclude that all the scaling laws and solutions are similar to those in MCT when we treat the multi-point density correlators at all GMCT levels equally.  However, we emphasize that all the important parameters including $a$, $b$, $\lambda$, $\gamma$, and $\sigma$ depend in a non-trivial manner on the closure level applied. These parameters 
are improved for higher levels $N$, as discussed in the accompanying paper.\cite{luo2019generalized} Finally, from our numerical tests based on the PY hard-sphere static structure factor, we find that the applicability ranges of the derived scaling laws do not differ much among different GMCT closure levels. 

Our work provides a solid mathematical analysis of the first-principles-based GMCT hierarchy for structural glass formers, extending the celebrated MCT scaling laws to dynamical multi-point density correlation functions. 
We add that the analysis is applicable to any kind of MF closure as long as  $\phi_N(k_1,\hdots,k_N,t)$ is closed by a linear combination of the product of other lower-level density correlation functions.  Table \ref{tab:coeff2} show some of the parameters for PY hard spheres under closure MF-$N[1^N]$, which are consistent with those in the accompanying paper.\cite{luo2019generalized} With the fast development of computational power, we also expect that the higher-order density correlation functions predicted by GMCT will become accessible in numerical simulations. This should provide a stringent test on the accuracy of the present GMCT framework and the here presented analytical asymptotic solutions. 

Finally, let us outline future directions of research to shed more light on glassy physics from a  first-principles perspective. Regarding the structural relaxation dynamics, we recall that the time- and wavenumber-dependent intermediate scattering functions predicted by GMCT for hard spheres are already in near-quantitative agreement with computer simulations, at least in the accessible simulation regime.\cite{janssen2015microscopic} However, note that the present GMCT framework invokes Gaussian and convolution approximations for all higher-order static correlations, i.e.\ all microstructural information is assumed to be contained in $S(k)$, which may severely affect the predicted dynamics of higher-order dynamic correlations. It is therefore possible that the predicted $\phi_2(k_1,k_2,t)$ will not agree with simulation results, but the predicted $\phi_1(k,t)$ accidentally does. Hence, it will be crucial to directly compare multi-point density correlation functions such as $\phi_2(k_1,k_2,t)$ obtained from simulation or experiment with the scaling laws derived in this paper. Similarly, the question to what extent the higher-order dynamical correlations predicted by GMCT can quantitatively capture the emergence of dynamical heterogeneities, and the related breakdown of the Stokes-Einstein relation, still remains to be explored. It will also be very interesting to study how static higher-order generalizations of $S(k)$, as well as their relation with e.g.\ locally preferred structural motifs\cite{robinson2019morphometric} and longer-ranged amorphous order metrics,\cite{zhang2020revealing} can be embraced into the theory. This development will be a crucial step toward the ultimate elucidation of the complex structure-dynamics link in glassy liquids.

\acknowledgments
We acknowledge the Netherlands Organisation for Scientific Research (NWO) for financial support through a START-UP grant.

\bibliography{paper2}

\end{document}